\documentclass[12pt,reqno]{amsart}
\usepackage{fullpage}

\usepackage{tikz}
\usetikzlibrary{matrix,arrows,decorations.pathmorphing}
\usepackage{tikz-cd}
\usetikzlibrary{arrows}

\usepackage{float}

\usepackage{amsmath}
\usepackage{mathrsfs}
\usepackage{soul}

\usepackage{epsfig}
\usepackage{amsmath,amsfonts,amsthm,amscd, amssymb}
\usepackage{latexsym}
\usepackage{enumerate}
\usepackage{multirow}
\usepackage{comment} 
 
\usepackage{chngcntr}
\makeatletter
\let\c@equation=\c@subsubsection

\makeatother
 
\newcommand*{\rom}[1]{\expandafter\@slowromancap\romannumeral #1@}
\newcommand{\Proof}[1]{\begin{proof} #1 \end{proof}}
\newtheorem{thm}{Theorem}[section]
\newtheorem*{theorem*}{Theorem}
\newtheorem*{question*}{Question}

\newcommand{\Thm}[1]{\begin{thm} #1 \end{thm}}
\newtheorem{pro}[thm]{Proposition}
\newcommand{\Pro}[1]{\begin{pro} #1\end{pro}}
\newtheorem{cor}[thm]{Corollary}
\newcommand{\Cor}[1]{\begin{cor} #1 \end{cor}}
\newtheorem{lem}[thm]{Lemma}
\newcommand{\Lem}[1]{\begin{lem} #1 \end{lem}}

\newtheorem{defn}[thm]{Definition}
\newcommand{\Def}[1]{\begin{defn} #1\end{defn}}
\newtheorem{rem}[thm]{Remark}
\newcommand{\Rem}[1]{\begin{rem} #1 \end{rem}}
\newtheorem{ex}[thm]{Example}
\newcommand{\Ex}[1]{\begin{ex} #1 \end{ex}}

\newtheorem{obs}[thm]{Observation}

\newtheorem{note}[thm]{Note}

\newtheorem{nota}[thm]{Notation}

\newcommand{\catC}{\mathbf{C}}
\newcommand{\catSet}{\mathbf{Set}}

\newcommand{\catAb}{\mathbf{Ab}}

\newcommand{\catTop}{\mathbf{Top}}

\newcommand{\catVect}{\mathbf{Vect}} 

\newcommand{\hocolim}[1]{\ensuremath{ \underset{#1}{\text{hocolim}}\; }}

\newcommand{\ilim}[2]{\ensuremath{\underset{\underset{#1}{\leftarrow}}{\text{lim}^{#2}}\;}}

\newcommand{\tightoverset}[2]{%
\mathop{#2}\limits^{\vbox to -.5ex{\kern-0.25ex\hbox{$#1$}\vss}}}

\newcommand{\Hom}{\text{Hom}}

\newcommand{\im}{\text{Im}}

\newcommand{\ev}{\text{ev}}

\newcommand{\op}{\text{op}}

\newcommand{\ch}{\text{ch}}

\newcommand{\Prin}{\text{Prin}}

\newcommand{\Tr}{\text{Tr}}

\newcommand{\Den}{\text{Den}}

\newcommand{\twobytwo}[4]{\ensuremath{\left( \begin{matrix}
#1 & #2 \\
#3 & #4
\end{matrix}  \right) }}

\newcommand{\set}[1]{\ensuremath{ \lbrace #1 \rbrace }}
\newcommand{\bra}[1]{\ensuremath{ | #1 \rangle }}

\newcommand{\Span}[1]{\ensuremath{ \langle #1 \rangle }}

\newcommand{\Cl}{\text{Cl}}

\newcommand{\com}{\text{com}}

\newcommand{\tr}{\text{tr}}

\newcommand{\dt}{\bullet}

\newcommand{\bi}{\mathfrak{b}}
\newcommand{\qu}{\mathfrak{q}}

\newcommand{\FF}{\mathbb{F}}
\newcommand{\ZZ}{\mathbb{Z}}

\newcommand{\RR}{\mathbb{R}}
\newcommand{\CC}{\mathbb{C}}

\newcommand{\jJ}{\mathcal{J}}

\newcommand{\iI}{\mathcal{I}}

\newcommand{\fF}{\mathcal{F}}
\newcommand{\pP}{\mathcal{P}}

\newcommand{\hH}{\mathcal{H}}

\newcommand{\mM}{\mathcal{M}}
\newcommand{\eE}{\mathcal{E}}

\newcommand{\cC}{\mathcal{C}}

\newcommand\Bcx{B_{\text{cx}}}
\newcommand\Princx{\text{Prin}_{\text{cx}} }

\usepackage[numbers,sort&compress]{natbib}
\title{Classifying space for  quantum  contextuality}
\date{\today}
\address{Department of Physics and Astronomy, University of British Columbia, Vancouver BC  V6T 1Z4}
\email{okay@math.ubc.ca/okay@phas.ubc.ca}

\address{Department of Mathematics, University of British Columbia, Vancouver BC  V6T 1Z4}
\email{dshein@math.ubc.ca}

\author{C\.{I}han Okay* and Daniel Sheinbaum}

\begin{document}
  \maketitle 
 
\begin{abstract} 
We construct a topological space to study contextuality in quantum mechanics. The resulting space is a classifying space in the sense of algebraic topology. Cohomological invariants of our space correspond to physical quantities relevant to the study of contextuality. Within this framework the Wigner function of a quantum state can be interpreted as a class in the twisted $K$-theory of the classifying space. 
\end{abstract}

\section{Introduction} 

In quantum mechanics only mutually commuting observables can be simultaneously measured to reveal their joint outcome. Such a set of commuting observables is called a context. It is a fundamental property of quantum mechanics that its predictions cannot be reproduced by assuming   pre-determined context independent outcomes---a phenomenon called contextuality {\color{black}  first observed in the no-go theorems of Bell \cite{bell64,bell1966problem} and Kochen--Specker \cite{kochen}}. This special feature of quantum systems is expected to play an important role in any computation scheme relying on quantum principles. Indeed, {\color{black}  contextuality has been} established as a resource to achieve universal quantum computation and quantum speed-up can only be attained if contextuality is present \cite{Howard14,Delfosse17,DBGR15,BDBOR17,AB09}.
To better characterize contextuality from the quantum computation perspective different frameworks have been developed such as  \cite{Cabello,Acin,Spekkens,abramsky,Coho}. 
{\color{black}
In this paper we extend the topological framework of \cite{Coho} based on chain complexes by introducing a topological space called the classifying space for contextuality. We  make a  connection to the sheaf-theoretic approach of \cite{abramsky,abramsky2011cohomology} based on \v Cech cohomology, which is applicable to a strong form of contextuality known as All-versus-Nothing contextuality \cite{abramsky2015contextuality}. For a recent attempt  at establishing a connection from the other direction see \cite{aasnaess2020cohomology}.
}

Classifying spaces are fundamental objects in algebraic topology that play a prominent role in the classification of principal bundles \cite{Milnor-classifying}. These geometric objects appear in gauge theories in the standard model \cite{Frankel-gauge}.
Motivated by questions  arising from gauge theory \cite{kac2000vacuum,witten1982constraints}, Adem et al. \cite{ACT12,AG15} introduced the notion of a classifying space for commutativity in Lie groups.  The classifying space for contextuality introduced in the present paper is a variant of this construction tailored for applications to contextuality. Not only is it a unifying object for contextual interpretations, it is {\color{black}also} an interesting mathematical object in its own right because of its  rich homotopy-theoretic structure.  The classifying space for commutativity of  extraspecial groups studied in \cite{O14,O15,O16} demonstrates such an interesting behavior.  
There is a close connection to quantum computation since   the basic observables called the Pauli observables constitute a finite group which is  an extraspecial group.
The goal of this paper is to make this connection precise by showing how traditional quantities relevant to contextuality (such as Wigner functions) can be viewed as elements in algebro-topological groups coming from our classifying space.

{\color{black}In this paper we consider Pauli observables that appear in the stabilizer formalism. In certain schemes of quantum computation, e.g. quantum computation with magic states \cite{magic} and measurement-based quantum computation \cite{MBQC}, these observables play an important role.} A context specifies a set of observables that can be measured simultaneously. 
In more mathematical terms for us  a context   refers to an abelian subgroup of the Pauli group. Given a collection $\iI$ of contexts 
we construct a topological space $\Bcx(\iI)$, called the classifying space for contextuality, and identify two classes:
\begin{itemize}
\item a cohomology class in the second cohomology group with mod-$p$ coefficients
$$
[\beta]\in H^2(\Bcx\iI)
$$ 

\item  a $K$-theory class, that depends on a quantum state $\rho$, in the $\beta$-twisted $K$-group   
$$
[W_\rho] \in \RR\otimes K^\beta(\Bcx\iI).
$$
\end{itemize}
The cohomology class $[\beta]$ is introduced in \cite{Coho} using chain complexes obtained from a cover of contexts. {\color{black} This chain complex coincides with the chain complex of $\Bcx\iI$ in degrees $\leq 3$.}
{\color{black} However,} the space itself is infinite-dimensional, {\color{black} meaning that there is no bound on the dimension of the cells and hence on the dimension of the chains,} analogously to the case of  classifying spaces of finite groups. 
{\color{black}The cohomology class $[\beta]$} captures {\color{black}state-independent} contextuality   in the {\color{black}Kochen--Specker sense, i.e.  detecting}   the  failure of assigning pre-determined measurement outcomes {\color{black} to all Pauli observables \cite{Coho}}. The full formulation of contextuality involves probability distributions on outcomes over contexts and can be formalized using sheaf theory \cite{abramsky}.
 Given a quantum state $\rho$ the empirical model $e_\rho$ provides a description as a probability distribution on measurement outcomes over each context. If the empirical model can not be described using a deterministic hidden-variable model then it is called contextual. When this approach is transferred to our topological language the empirical model $e_\rho$ can equivalently be described as the class $[W_\rho]$ which lives in the twisted   $K$-group of $\Bcx\iI$. This class is essentially the discrete Wigner function \cite{Gro06} of the quantum state.   
 
The structure of the paper is as follows. In \S \ref{sec:contextuality} we start with the sheaf-theoretic definition of contextuality, and restrict this framework to Pauli observables. This is summarized in Corollary \ref{cor:contextEbeta}. Section \ref{sec:Wigner} uses twisted representations to describe the space of empirical models. Theorem \ref{thm:Wigner} explains the relationship between the empirical model $e_\rho$ and the class $[W_\rho]$ corresponding to the Wigner function. The classifying space for contextuality is introduced in \S \ref{sec:classifying}. We study its cohomology and explain the distinction between the even and odd prime cases.   In particular, for $p=2$ we give a cup product decomposition for $[\beta]$ (Proposition \ref{pro:cup}). We refer to the quantum computation literature to point out the consequences of these observations. We summarize homotopy-theoretic properties of $\Bcx\iI$ that are mainly obtained in \cite{O16}. The representation-theoretic approach of \S  \ref{sec:Wigner} and the topological approach of \S \ref{sec:classifying} come together in \S \ref{sec:K-theory}. In this section we compute the $\beta$-twisted   $K$-theory of $\Bcx\iI$ in  Theorem \ref{thm:K-theory-compute} whose proof mostly follows the  computation in \cite{O14} for the untwisted version. We discuss some examples in \S \ref{sec:examples}. First, the examples of the Mermin square
and the Mermin star are discussed in the current topological framework. Then we give a generalized version of the {\color{black}Mermin inequality} from a representation-theoretic point of view.

 \subsection*{Acknowledgement} The authors would like to thank Robert Raussendorf for useful discussions. This work is supported by NSERC (CO), the Stewart Blusson
Quantum Matter Institute (CO) and  CONACYT 291928 (DS).

\section{Contextuality for Pauli observables}\label{sec:contextuality}


{\color{black}
In this section we introduce the notion of quantum contextuality in the sheaf-theoretic framework of  \cite{abramsky} for the group of Pauli observables.}
Our description makes a connection to the topological framework of \cite{Coho}. For basics of quantum computation we refer to \cite{Nielsen02}.

\subsection{Contexts in quantum mechanics} {\color{black}In quantum computation the Hilbert space $\hH$ is finite-dimensional.} In this paper we will consider the case when the dimension $d$ is a power of a prime $p$.  The state of a system is specified by a density matrix $\rho$, {\color{black} a Hermitian matrix acting on   $\hH$ that is positive semi-definite and has trace $\Tr(\rho)=1$.} 
 We will denote the space of density matrices by $\Den(\hH)$.  
 When $p=2$ the observables are Hermitian matrices as usual, but for $p>2$ the type of observables we will consider are unitary matrices that are not necessarily Hermitian. {\color{black}
In this formulation the set of observables can be given the structure of a group. This is more convenient for our purposes since  we are interested in the algebraic relations among the observables. 
} In fact, the results of this paper can be applied to the more general observables introduced in \cite{Coho}. But we restrict to Pauli observables since for this case the topology of the classifying space for contextuality is well-understood, and independently studied (see Theorem \ref{thm:homotopy-groups}).

An observable $A$ can be measured on a system at state $\rho$. The measurement result is specified by an eigenvalue $\lambda_A$ of the observable.
 Born's rule says that the probability of a measurement result  $\lambda_A$ occurring is given by  the trace $\Tr(\rho P_{\lambda_A} )$ where $P_{\lambda_A}$ is the projector onto the eigenspace corresponding to the eigenvalue $\lambda_A$.
This can be generalized to a set of pairwise commuting observables, also known as  a {\it context}.  Let $C$ be a context and $\lambda$ be a function that assigns an eigenvalue to each observable in the context. 
{\color{black}The probability of, when measuring all the observables
in $C$ simultaneously,
obtaining the outcome $\lambda_A$ for the measurement $A$ for all $A\in C$
is given by  $\Tr(\rho P_{\lambda})$ where $P_{\lambda}$ projects onto the common eigenspace of the observables in $C$ which
corresponds to the eigenvalues $\set{\lambda_A}_{A\in C}$.}

\subsection{{\color{black}(Pre)}sheaf-theoretic description of contextuality}  

Abramsky and Brandenburger introduced a sheaf-theoretic approach in \cite{abramsky} to study probability distributions for  measurements over various contexts. 
The general framework consists of
\begin{itemize}
\item a set $X$ of measurements,
\item a set $O$ of outcomes,
\item a collection $\mM$ of contexts such that
$$
X= \bigcup_{C\in \mM} C.
$$
\end{itemize}
The collection $\mM$ is  sometimes referred to as a \textit{cover of contexts}.
{\color{black}Following \cite{abramsky2019comonadic}} we will assume that $\mM$ is closed under intersections and regard it as a partially ordered set (poset)  under inclusion. {\color{black}For a set $U$ let $\eE(U)$ denote the set of functions $U\to O$.}
Given this data one can construct a {\color{black}{\it presheaf of events}\footnote{{\color{black}In \cite{abramsky} the presheaf of events, which is in fact a sheaf, is defined as a functor $\eE:\pP(X)^\op \to \catSet$, where $\pP(X)$ denotes the poset of all subsets of $X$.   We prefer to work with the presheaf $\eE:\mM^\op\to \catSet$ in order to make a connection to twisted $K$-theory in \S \ref{sec:K-theory}. } }}
$$\eE:\mM^\op \to \catSet$$
which assigns the set $\eE(C)$ to a given context $C$.  
Let $D(U)$ denote the set of probability distributions over $U$ with finite support. Consider the functor
$$D:\catSet \to \catSet$$
where for a function $f:U\to U'$ the corresponding function $D(f):D(U)\to D(U')$ is defined by
$$
D(f)(d):U'\to \RR_{\geq 0},\;\;\; u'\mapsto \sum_{u\in f^{-1}(u')} d(u). 
$$ 
We are interested in the inverse limit of the composite functor $D\eE:\mM^\op\to \catSet$. 
{\color{black} An element   of the inverse limit is a collection of distributions  $d=\set{d_C}_{C\in \mM}$, where $d_C$ belongs to $D\eE(C)$, such that further restriction to pairwise  intersections coincide: $(d_C)|_{C\cap C'} = (d_{C'})|_{C'\cap C}$.} Thus the inverse limit is a subset of the product 
$$
\ilim{}{}D\eE \subset \prod_{C\in \mM} D\eE(C).
$$
In physics literature an element of the inverse limit is called an {\it empirical model}.

An important example comes from quantum mechanics. Given a state $\rho$ we can define an element $e_\rho$ of the inverse limit. Let us denote this assignment by a map 
$$
e:\Den(\hH) \to \ilim{}{}D\eE,\;\;\;\;\rho\mapsto e_\rho
$$
where   $(e_\rho)_C$ is defined on  $\lambda:C\to O$ by the quantum mechanical probability
$$
(e_\rho)_C(\lambda) = \Tr(\rho P_\lambda) \in \RR_{\geq 0}.
$$
Let $S(\rho)$ denote the set of functions $s\in \eE(X)$ such that {\color{black}the empirical model $e_\rho$ satisfies  $ (e_\rho)_C(s|_C)\neq 0$} for all $C\in\mM $.

\Def{\rm{\label{def:contextual} A state  $\rho$ is called {\it non-contextual} if there exists a distribution $d'\in D\eE(X)$ such that $d'|_C=(e_\rho)_C$ for all $C\in \mM$, otherwise, it is called {\it contextual} (with respect to $\mM$). A state $\rho$ is called {\it strongly contextual} if the set $S(\rho)$ is empty.
}}

In the language of {\color{black}(pre)}sheaf theory  we can summarize this definition by the statement that a state $\rho$ is contextual if  $e_\rho$ does not come from a global section, also known as a deterministic hidden-variable model.

\subsection{Pauli group} {\color{black} In this paper we will be interested in observables that appear in the stabilizer formalism, a subtheory of quantum mechanics that plays an important role in certain computation schemes  such as quantum computation with magic states \cite{magic} and measurement-based quantum computation \cite{MBQC}.}
{\color{black}In this formalism the set of observables consists of   the Pauli observables}
 acting on a Hilbert space $\hH$ of dimension $d$. More precisely, $\hH$ is  the complex group ring $\CC[\ZZ/d]$ of the additive group $\ZZ/d$ of integers modulo $d$.
{\color{black} The shift $X$ and   phase $Z$ operators are defined by }
\begin{equation*}\label{XZ}
Z\bra{a}=\omega^{a}\bra{a} \;\;\;\; X\bra{a} =\bra{a+1}
\end{equation*}
where $a\in \ZZ/d$ and $\omega$ is the first $d$--th root of unity $e^{2\pi i/d}$. We are using the quantum mechanical notation $\bra{a}$ to denote the basis elements of the Hilbert space. This structure  is usually called a $1$-qudit, or more popularly a $1$-qubit when $d=2$.

Qudits can be composed to obtain larger systems. An $n$-qudit has the Hilbert space
$$
\hH_n = \underbrace{\hH\otimes \cdots \otimes \hH}_n
$$ 
which can be identified with the complex group ring $\CC[(\ZZ/d)^n]$.    
For $a\in (\ZZ/d)^n$ we introduce the notation
$$
Z(a) = Z^{a_1} \otimes \cdots \otimes Z^{a_n}  \;\;\;\;
X(a)= X^{a_1} \otimes \cdots \otimes X^{a_n}.
$$
Throughout the paper we will only consider the case where $d$ is a prime $p$.
When it comes to defining the Pauli group, or the group of Pauli observables, there is a distinction between the cases $p=2$ and $p>2$. Let us set 
$$
\mu= \left\lbrace \begin{array}{ll}
\sqrt{\omega} & \text{ if } p=2\\
\omega & \text{ if } p>2.
\end{array}
\right.
$$
{\color{black}Let $I_{p^n}$ denote the $p^n\times p^n$ identity matrix.}
The Pauli group of an $n$-qudit system is defined by {\color{black}
\begin{equation}\label{eq:pauli-def}
P_n=\Span{\mu I_{p^n}, Z(a),X(a)|\; a\in (\ZZ/p)^n} 
\end{equation} }
as a subgroup of the unitary group $U(p^n)$.   
 
\Ex{\rm{The Hilbert space of a $1$-qubit system is given by $\hH = \CC[\ZZ/2]$. Pauli observables constitute a subgroup of the unitary group $U(2)$ generated by the Pauli matrices
$$
X= \twobytwo{0}{1}{1}{0}\;\;\;\; Z= \twobytwo{1}{0}{0}{-1}
$$ 
together with $iI_2$.
}}

\subsection{Extension class} 
As an abstract group $P_n$ is an extraspecial $p$-group. For properties of extraspecial $p$-groups see \cite{As86}. When $p=2$ it is of complex type, and for $p>2$ it has exponent $p$.


{\color{black}Let us set $V=(\ZZ/p)^{n} \times (\ZZ/p)^{n} $.}
The subgroup $\Span{\mu I_{p^n}}\subset P_n$ 
is isomorphic to the additive group 
$$
Z_\mu = \left\lbrace
\begin{array}{cc}
\ZZ/4 & p=2 \\
\ZZ/p & p>2 
\end{array}
\right.
$$
under the canonical homomorphism {\color{black}$\mu I_{p^n}\mapsto \mu$}.
 Under this identification the Pauli group can be written as a  central extension 
\begin{equation}\label{eq:extension}
0\to Z_\mu \to P_n \stackrel{\pi}{\longrightarrow} V \to 0
\end{equation}
since up to an element of $\Span{\mu I_{p^n}}$ a Pauli operator is specified by an element of $V$. {\color{black}The standard procedure \cite[Chapter 6]{weibel1995introduction} in the theory of group extensions is to choose a set-theoretic section of $\pi$ which can be used to calculate the cocycle describing the extension. This cocycle is usually referred to as the extension cocycle.  } Given a function $\gamma:V\to Z_\mu$ we can 
{\color{black}define} a set-theoretic section {\color{black} $\eta_\gamma:V\to P_n$  by setting}
$$
\eta_\gamma(v) =\mu^{\gamma(v)} Z(v_z)X(v_x), {\color{black}\;\;\; v=(v_z,v_x)\in V.}
$$
 The extension cocycle can be computed from the formula {\color{black}
\begin{equation}\label{eq:extcocycle}
\mu^{\beta_\gamma(v,v')} = \eta_\gamma(v)\eta_\gamma(v')\eta_\gamma(v+v')^{-1}.
\end{equation} 
 According to the classification of group extensions via group cohomology, the set of  equivalence classes of extensions of $V$ by $Z_\mu$ is in bijective correspondence with the elements of the cohomology group  $H^2(V,Z_\mu)$. These elements are equivalence classes of cocycles and they are referred to as cohomology classes of degree $2$.}
The cohomology class $[\beta_\gamma]\in H^2(V,Z_\mu)$ is independent of the choice of the section, {\color{black} and   corresponds to the group extension equivalent to the Pauli group $P_n$.}
    
Let $\eta_0$ denote the section corresponding to the zero function $\gamma=0$, then {\color{black}using \ref{eq:extcocycle} the corresponding cocycle can be computed as  
\begin{equation*} 
\begin{aligned} 
\mu^{\beta_0(v,v')}  &=  \eta_0(v)\eta_0(v')\eta_0(v+v')^{-1} \\
&=Z(v_z)X(v_x) Z(v'_z)X(v'_x) \left( Z(v_z+v_z')X(v_x+v_x') \right)^{-1}\\
&=    \omega^{v_x\cdot v_z'}
\end{aligned} 
\end{equation*}} 
where the notation $v\cdot w$ stands for the standard inner product.
Therefore we find that 
\begin{equation}\label{eq:beta0}
\beta_0(v,v')  = \left\lbrace 
\begin{array}{cc} 
v_x\cdot v_z'  &  p>2 \\
 2\,v_x\cdot v_z'  &  p=2.
\end{array}
\right.
\end{equation} 
  {\color{black} For a  given function $\gamma:V\to Z_\mu$ let us define  $d\gamma: V\times V\to Z_\mu$, the coboundary of $\gamma$, by the formula $d\gamma(v,v')=\gamma(v)-\gamma(v+v')+\gamma(v') $.}
For any other choice of $\gamma$ we have that $\beta_\gamma = \beta_0 +d\gamma$.

\subsection{Isotropic subspaces} {\color{black}It is convenient to identify $\ZZ/p$ with the finite field $\FF_p$, and regard $V$ as a vector space over $\FF_p$. }
Commutation properties of Pauli operators are determined by a symplectic form {\color{black} $\bi:V\times V\to \ZZ/p$. This form can be derived as follows
$$
\begin{aligned}
\eta_\gamma(v)\eta_\gamma(v')\left( \eta_\gamma(v')\eta_\gamma(v) \right)^{-1} &=  \mu^{\beta_\gamma(v,v')}\eta_\gamma(v+v')\mu^{ - \beta_\gamma(v',v)}\eta_\gamma(v+v')^{-1} \\
&= \omega^{v_x\cdot v_z' - v_x'\cdot v_z}.
\end{aligned}
$$
Therefore, identifying $\Span{\omega}$ with the additive group $\ZZ/p$ via the assignment $\omega \mapsto 1$, we obtain}
\begin{equation} \label{eq:bi}
\bi(v,v') =  v_x\cdot v_z' - v_x'\cdot v_z.
\end{equation}
We can  choose a symplectic basis $\set{z_1,\cdots,z_n,x_1,\cdots,x_n}$ for $V$ where $x_i$ and $z_i$ are the symplectic pairs satisfying $\bi(z_i,x_i)=1$. The suitable choice 
is to take $z_i$ the canonical basis for the first factor of $V=(\ZZ/p)^n\times (\ZZ/p)^n$ and $x_i$ the basis for the second factor.

\Def{{\rm
A subspace $I\subset V$ is called {\it isotropic} if  $\bi|_I=0$  i.e. $\bi(v,v')=0$ for all $v,v'\in I$.  We write $\iI(V)$ for the collection of all isotropic subspaces in $V$.
}}

We will derive a property of $\beta$   restricted onto an isotropic subspace. For this purpose we make a canonical choice for the extension cocycle:{\color{black}
\begin{equation}\label{eq:beta-canonical}
\beta = \beta_0 + d\bar\qu
\end{equation}
where $\bar\qu:V\to Z_\mu$ denotes the function
$$
\bar\qu(v) = \left\lbrace 
\begin{array}{cc}
 v_x\cdot v_z & p=2\\
(v_x\cdot v_z)/2 & p>2.
\end{array}
\right.
$$
Note that $2$ is invertible in $\ZZ/p$ when $p>2$. For the rest of the paper we use  $\eta_{\bar\qu}$ as our set-theoretic section, and denote it simply as  $\eta$.}
The motivation for this choice is provided by the following result.

\Pro{\label{pro:beta-canonical}
For any isotropic subspace $I$ the restricted cocycle satisfies $2\beta|_I=0$. Moreover, for $p>2$ we have that $2\beta=\bi$.
}
\Proof{For $p=2$ as a consequence of \ref{eq:beta0} we have that {\color{black}$2\beta=2d\bar\qu=2(v_x\cdot v_z' + v_x'\cdot v_z)\mod 4$.} Thus $2\beta$ restricted to $I$ vanishes. In the case of $p>2$  one can calculate to obtain $\beta=\bi/2$.
}

\subsection{Contextuality for Pauli observables} 
{\color{black} We will consider the Pauli observables whose eigenvalues are given by the set  $\set{1,\omega,\cdots,\omega^{p-1}}$. }
These observables are precisely the elements of order $p$ in the Pauli group{\color{black}, and they play an important role in quantum contextuality \cite{Coho}.}

In the {\color{black}(pre)}sheaf-theoretic description of contextuality for Pauli observables we consider the triple $(\Sigma,\iI,\ZZ/p)$ where 
\begin{itemize}
\item  the collection  $\iI$ of contexts  is given by a subcollection of $\iI(V)$ satisfying the property that if $I,I'\in\iI$ then $I\cap I'\in \iI$

\item the set of measurements is  $\set{\eta(a)|\;a\in \Sigma(\iI)}$ where
$$
\Sigma(\iI) = \bigcup_{I\in \iI} I
$$

\item the set of outcomes is $\ZZ/p$.
\end{itemize}

\Rem{\label{rem:betaI}{\rm
As a consequence of Proposition \ref{pro:beta-canonical} the cocycle $\beta|_I$ is the zero function when $p>2$, and is divisible by $2$ when $p=2$. For the $p=2$ case we identify $2(\ZZ/4)$ with $\ZZ/2$ and regard $\beta|_I$ as a function $I\times I\to \ZZ/2$.
}}

Next we introduce a modified version of the event {\color{black}presheaf by focusing on the support of the empirical models associated to quantum states}. 
In the Pauli case we will use the set of functions
$$
\eE_\beta(I) = \set{s:I\to\ZZ/p|\; \beta|_I=ds}
$$ 
which  gives a functor
$$
\eE_\beta:\iI^\op \to \catSet.
$$
Note that this is a subfunctor of the {\color{black}presheaf $\eE$} of events. The motivation for this definition is the following observation in quantum mechanics (see also \cite{Delfosse17}).

\Pro{\label{pro:Ebeta}If a function $s:I\to \ZZ/p$ does not belong to $\eE_\beta(I)$ then $(e_\rho)_I(s)=0$ {\color{black}for all quantum states $\rho$}.}

\Proof{If $(e_\rho)_I(s)\neq 0$ then there is a common eigenstate $\bra{s}$ with eigenvalue $\omega^{s(a)}$ for the observable $\eta(a)$. Thus we can write
$$
\omega^{s(a)+s(b)}\bra{s} = \eta(a)\eta(b)\bra{s} = \omega^{\beta(a,b)} \eta(a+b)\bra{s} =  \omega^{\beta(a,b)+s(a+b)} \bra{s} 
$$
which implies that $s\in \eE_\beta(I)$.
}

This means that the distribution $(e_\rho)_I \in D\eE(I)$ can be regarded as an element of $D\eE_\beta(I)$ for each $I\in \iI$. 
As a consequence the empirical model $e_\rho$ associated to a state $\rho$ 
gives a function
$$
e:\Den(\hH) \to \ilim{}{} D\eE_\beta.
$$

Proposition \ref{pro:Ebeta} has another implication {\color{black}which will be useful in the next sections. To demonstrate this implication we first need to introduce the set}
$$
\eE_\beta(\iI) = \ilim{}{}\eE_\beta
$$
as the inverse limit of the functor $\eE_\beta:\iI^\op\to \catSet$. More explicitly, this set consists of functions that are obtained by``gluing" functions $\eE_\beta(I)$ over contexts.
We have a commutative diagram
\begin{equation}\label{diag:DenLim}
\begin{tikzcd}
\Den(\hH) \arrow{r}{e} \arrow{rd} & 
\ilim{}{}D\eE_\beta \arrow{d} & D\eE_\beta(\iI) \arrow[l,"\theta"'] \arrow{d} \\
& \ilim{}{} D\eE & D\eE(\Sigma) \arrow{l}
\end{tikzcd} 
\end{equation}
where the union of all contexts is denoted by $\Sigma=\Sigma(\iI)$.

{\color{black} We reformulate contextuality using the subfunctor $\eE_\beta$ and the map $\theta$ in  \ref{diag:DenLim} to be used in Section \ref{sec:Wigner}.}

\Cor{\label{cor:contextEbeta} 
Let $\rho$ be a state and $\iI$ be a cover of contexts.
\begin{enumerate}
\item $\rho$ is strongly contextual if $\eE_\beta(\iI)=\emptyset$.

\item  $\rho$ is contextual if and only if $e_\rho \notin \im(\theta)$.
\end{enumerate}
}
\Proof{
Recall from Definition \ref{def:contextual} that $\rho$ is strongly contextual if  $S(\rho)$ is empty. By definition $s\in\eE(\Sigma)$ belongs to this set if $(e_\rho)_I(s|_I)\neq 0$ for all $I\in \iI$. By Proposition \ref{pro:Ebeta} the restriction $(e_\rho)_I$ belongs to $\eE_\beta(I)$ for each $I$. This implies that $s|_I$ is in $\eE_\beta(I)$. But this suffices to conclude that $s$ belongs to $\eE_\beta(\iI)$, since it is obtained by patching the local functions.

For the second statement observe that
 if $e_\rho$ can be written as $\theta(d')$ for some $d'\in D\eE_\beta(\iI)$ then regarding $d'$
 as an element of $D\eE(\Sigma)$ in the obvious way implies that $\rho$ is non-contextual.  Conversely, it suffices  to check that if $e_\rho = \theta(d'')$ for some $d''\in D\eE(\Sigma)$ then $d''$ belongs to $D S(\rho) $, which is a subset of $D\eE_\beta(\iI)$ by the previous paragraph. Observe that if $d''(s)\neq 0$ then 
$
(e_\rho)_I(s|_I)= \theta(d'')_I(s|_I) \neq 0 
$
for all $I$, and thus $s$     belongs to $S(\rho)$. 
}
 
{\color{black} In physics literature the situation $\eE_\beta(\iI)=\emptyset$ is referred to as {\it state-independent contextuality}, a version of strong contextuality independent of the chosen state. }

\section{Wigner function as a global section}
\label{sec:Wigner}

The purpose of this section is to interpret empirical models described in the previous section in representation-theoretic terms. The approach taken here is to extend probability distributions over $\RR_{\geq 0}$ to the whole field of real numbers. {\color{black}The relationship between negative probabilities and empirical models (no-signaling models) was previously studied in \cite{abramsky,abramsky2014operational}. Our goal is to understand the inverse limit describing the space of empirical models.}  After extending the coefficients the empirical model map {\color{black}in \ref{diag:DenLim} becomes
$$
e: \Den(\hH) \to \ilim{}{} {\color{black}\RR \eE_\beta}
$$
where $\RR\eE_\beta:\iI^\op\to \catVect_\RR$ is the functor that sends an isotropic subspace $I$ to the $\RR$-vector space $\RR\eE_\beta(I)$ generated by the set   $\eE_\beta(I)$.}
Our main result {\color{black}(Proposition \ref{pro:inv-lim-rep})} is the identification of the inverse limit {\color{black}in the target of $e$} as the representation ring $\RR\otimes R(V)$. Under this identification the empirical model $e_\rho$ corresponds to the Wigner function $W_\rho$ of the state $\rho$. In general $W_\rho$ is not a probability distribution and may assume negative values. An important application to contextuality, for $p>2$, says that   $W_\rho$ is a probability distribution  if and only if $\rho$ is non-contextual. This result displays a distinct feature of the odd prime case as first demonstrated in \cite{Delfosse17,Howard14}.

\subsection{Twisted representations} {\color{black}Let $I\subset V$ denote an isotropic subspace. We will regard $I$ as an additive group. Recall from Remark \ref{rem:betaI} that restriction of $\beta$ gives a $2$-cocycle $\beta|_I:I\times I\to \ZZ/p$ for all $p$ (it is zero when $p>2$ as a consequence of Proposition \ref{pro:beta-canonical}). For the sake of simplicity we will denote the restricted cocycle by $\beta$ as well.  There is a  corresponding cohomology class $[\beta]$ in $H^2(I,\ZZ/p)$, the group cohomology of $I$ with coefficients in $\ZZ/p$. }
Let $\iota:\ZZ/p\to U(1)$ denote the embedding $\iota(1)=\omega$ where $\omega$ is the $p$-th root of unity $e^{2\pi i/p}$. Under this identification the cohomology class $\beta$ corresponds to a class in {\color{black} $H^2(I,U(1))\cong H^2(I,\ZZ/p)$, also denoted by the same symbol. Here $H^2(I,U(1))$ denotes the group cohomology of $I$ with coefficients in $U(1)$ (as a trivial $I$-module).} Such a class can be used to define a  twisted representation \cite{serre1977linear}. 

In general, for a finite group $G$ and a cohomology class $\alpha\in H^2(G,U(1))$ an $\alpha$-twisted representation can be thought of as a  linear representation of the extension  $\tilde G$ corresponding to $\alpha$. Therefore most of the properties of linear representations can be transferred to twisted representations. 
There is a Grothendieck group of twisted representations \cite{dwyer2008twisted}, denoted by $R_\alpha(G)$, {\color{black}called the $\alpha$-twisted representation group. Note that for twisted representations tensor product of representations does not give a ring structure on $R_\alpha(G)$. For $[\alpha]=0$ we write $R(G)$ for the representation group, which is in fact a ring under tensor product.}

\Pro{\label{pro:rep-ring}The embedding $\iota:\ZZ/p\to U(1)$ induces a natural {\color{black} isomorphism of functors $\iota_*:\RR\eE_\beta\to \RR\otimes R_\beta$ whose component at an isotropic subspace $I\in \iI$ is given by the   isomorphism of $\RR$-vector spaces
$$
\iota_*(I): \RR \eE_\beta(I) \to \RR\otimes R_\beta(I)
$$
 defined by sending a basis element $s:I\to \ZZ/p$ to the twisted representation given by the composition $\chi_s: I\xrightarrow{s} \ZZ/p \xrightarrow{\iota} U(1)$.
}
} 

\Rem{\label{rem:untwist}{\rm
Note that when $p>2$ the twisted representation group coincides with the ordinary representation {\color{black}ring} $R(I)$ as a consequence of Proposition \ref{pro:beta-canonical}.
}}

{\color{black}
For a finite group $G$ we will recall some notions  from \cite{Chu15} related to $\alpha$-twisted representations where $\alpha$ is a $2$-cocycle $G\times G\to U(1)$. 
Let $\Cl_\alpha(G)$ denote the $\CC$-vector space   of $\alpha$-class functions. A function $f:G\to \CC$ is an $\alpha$-\textit{class function} if it satisfies the formula 
\begin{equation}\label{twisted}
f(hgh^{-1})=\frac{\alpha(h,h^{-1})}{\alpha(h,gh^{-1})\alpha(g,h^{-1})} f(g)
\end{equation}
for all $g,h\in G$. 
Consider   the character map:
\begin{equation}\label{eq:character}
{\color{black}\ch(G):}R_\alpha(G) \to \Cl_\alpha(G) 
\end{equation}
defined by sending a twisted representation $\pi:G\to U(m)$ to the associated character  $f_\pi:G\to \CC$ where $f_\pi(g)= \Tr(\pi(g))$.}

\Lem{\label{lem:set-of-functions} Let $I\subset V$ be an isotropic subspace. Then $\Cl_\beta(I)$ coincides with the $\CC$-vector space  $\CC^I$ of complex valued functions on $I$.}
\Proof{ 
The $p>2$ case follows from Remark \ref{rem:untwist}. For $p=2$ we check that 
$$
\frac{\omega^{\beta(w,-w)}}{\omega^{\beta(w,v-w)}\omega^{\beta(v,-w)}} = \frac{\eta(w)\eta(w)\eta(0)}{\eta(w)\eta(v+w)\eta(v)\, \eta(v)\eta(w)\eta(v+w) }= 1
$$
for all $v,w\in I$.   Recall that each $\eta(v)$ squares to the identity matrix, {\color{black} $\eta(0)=I_{2^n}$, and $[\eta(v),\eta(w)]=I_{2^n}$} since $v,w$ belong to an isotropic subspace. 
}
 
\subsection{Computing the inverse limit} Let $\iI\subset \iI(V)$ be a cover of contexts. We are interested in the inverse limit of the functor $\RR\eE_\beta : \iI^{\op} \to {\color{black}\catVect_\RR}$. We will compute it using the character map \ref{eq:character}. The first step is to extend the scalars to complex numbers.

\Lem{\label{lem:invlimit} There is a natural isomorphism of $\CC$-vector spaces
$$
\CC\otimes ( \ilim{}{} \RR \eE_\beta) \cong \CC^{\Sigma}.
$$
}
\Proof{Up to isomorphism we can consider the functor $\RR\otimes R_\beta$ instead of $\RR\eE_\beta$ (Proposition \ref{pro:rep-ring}). Since the character map \ref{eq:character} is natural it gives a natural transformation  $\ch:\RR\otimes R_\beta \to \Cl_\beta$, where $\Cl_\beta$ is regarded as a functor on $\iI^{\op}$. Tensoring $\RR\otimes R_\beta$ with $\CC$ the character map becomes an isomorphism of $\CC$-vector spaces. We  denote the resulting natural isomorphism by $\ch:\CC\otimes R_\beta \to \Cl_\beta$. Since taking inverse limit commutes with tensoring with $\CC$ we have
$$
\CC\otimes (\ilim{}{} \RR \otimes R_\beta) \cong \ilim{}{} \CC \otimes R_\beta \cong \ilim{}{} \Cl_\beta.
$$
Using Lemma \ref{lem:set-of-functions} we can compute the inverse limit of $\Cl_\beta$. The functor $\Cl_\beta$ is isomorphic to the functor  $\CC^-$ which sends $I$ to the set $\CC^{I}$ of complex valued  functions. Inverse limit of $\CC^-$  {\color{black}in $\catVect_\CC$} is simply the set of functions from the union $\Sigma=\Sigma(\iI)$  to   complex numbers {\color{black} regarded as a $\CC$-vector space in the natural way}. 
}

The next step is to identify the inverse limit over $\RR$. We take $\iI=\iI(V)$ and $\Sigma=V$ since smaller sets of measurements can be dealt with by naturality, i.e. restricting from $\iI(V)$. 
 The analysis depends on the prime $p$. Let us define a map
$$
\phi_p: \RR\otimes R(V) \to \ilim{}{} \RR\otimes R_\beta
$$
as follows:
\begin{itemize}
\item $p=2$: A $\beta$-twisted representation is a linear representation (compatible with the embedding $\iota$) of the extension
$$
0\to \ZZ/2 \to \tilde I \to I \to 0
$$
corresponding to the cocycle $\beta$. By definition of $\beta$ (\ref{eq:beta-canonical}) every element of the group  $\tilde I$ has order at most $2$. Therefore the class function of a representation takes values in $\RR$. This means that the character map \ref{eq:character} factors through the isomorphism
$$
{\color{black}\ch_\RR(I)}: \RR\otimes R_\beta(I) \to \RR^I.
$$
A similar observation is true for $\RR\otimes R(V)$, i.e. the character map induces an isomorphism to $\RR^V$. Using these identifications we can define
$$
\phi_2: \RR\otimes R(V) \to \ilim{}{} \RR\otimes R_\beta
$$
by sending a representation to the restriction of the associated character to each isotropic subspace {\color{black}and using the inverse of the isomorphism $\ch_\RR(I)$, i.e. $\pi:V\to U(1)$ is sent to $\set{\ch_\RR(I)^{-1}(f_{\pi}|_I)}_{I\in \iI}$.}
More explicitly, we have a commutative diagram 
$$
\begin{tikzcd}
\RR\otimes R(V) \arrow{r}{\phi_2} \arrow[d,"\ch_\RR(V)"'] & \ilim{}{} \RR\otimes R_\beta \dar[shorten <= -1.9ex ]  {\ilim{}{}\ch_\RR} \\
 \RR^V \arrow{r}{\cong}  & \ilim{}{} \RR^-
\end{tikzcd}
$$
 where $\RR^-$ is the functor $I\mapsto \RR^I$. The lower horizontal map is an isomorphism by an analogous argument to that given in the proof of Lemma \ref{lem:invlimit}. The point is that for $p=2$ extending coefficients to real numbers is sufficient.
\vspace{5mm}

\item $p>2$: We have seen that a $\beta$-twisted representation is the same as an ordinary representation. Therefore for each isotropic subspace we have a restriction map $R(V)\to R(I)$. These maps induce a map to the inverse limit
$$
\phi_p:\RR\otimes R(V) \to  \ilim{}{} \RR\otimes R_\beta
$$ 
and again this map turns out to be an isomorphism as in the $p=2$ case. This can be seen by extending the scalars to complex numbers and using Lemma \ref{lem:invlimit}. Indeed, tensoring $\phi_p$ with $\CC$ gives an isomorphism since we obtain a commutative diagram  
$$
\begin{tikzcd}
\CC\otimes R(V) \arrow{r}{\CC\otimes\phi_p} \arrow[d,"\ch(V)"'] &[1em]  \CC\otimes (\ilim{}{} \RR\otimes R_\beta)  \arrow{r}{\cong} & \ilim{}{} \CC\otimes R_\beta  \dar[shorten <= -1.9ex ]   {\ilim{}{}\ch}\\ 
 \CC^V \arrow{rr}{\cong}  && \ilim{}{} \CC^-
\end{tikzcd}
$$
Here left-vertical map is an isomorphism by definition of class functions and right-vertical arrow is an isomorphism by Lemma \ref{lem:invlimit}.
\end{itemize}

Both diagrams imply that $\phi_p$ is an isomorphism, and under the identification of Proposition \ref{pro:rep-ring} we obtain:  

\Pro{\label{pro:inv-lim-rep}  
There is a natural isomorphism
$$
\phi_p: \RR\otimes R(V) \to \ilim{\iI(V)}{} \RR \eE_\beta\;\;\;\; \text{ for any prime $p$. }
$$
} 
{\color{black}
\begin{rem}
{\rm
Theorem 5.9 in \cite{abramsky} characterizes empirical models  as probability models that admit a global section with  negative probabilities. Proposition \ref{pro:inv-lim-rep} can be used to deduce this theorem in the case of Pauli observables and empirical models represented by an element in the inverse limit of $D\eE_\beta$.
 More precisely, such an empirical model can be represented by a global section in $\RR\otimes R(V)$ given by the preimage of the empirical model under $\phi_p$.
Conversely, a global probability distribution represented by an element in $\RR\otimes R(V)$, i.e. probabilistic mixture of representations, gives an empirical model via restriction. 
}
\end{rem}
}
\Rem{\label{rem:free-abelian} \rm{
Note that the inverse limit of $R_\beta$ is a subgroup of the product $\prod_{I\in\iI(V)} R_\beta(I)$. In particular, it is a torsion-free abelian group. Therefore Proposition \ref{pro:inv-lim-rep} shows that the inverse limit of $R_\beta$ is isomorphic to $\ZZ^{|V|}$---the free abelian group of rank $|V|=p^{2n}$. 
}}

\subsection{Wigner functions} Density matrices map to the inverse limit of $ \RR \eE_\beta$ by the assignment $\rho\mapsto e_\rho$. As a consequence of the isomorphism $\phi_p$ of  Proposition \ref{pro:inv-lim-rep} there must be a map $W$ that fits in the commutative diagram 
\begin{equation}\label{diagram:Wigner}
\begin{tikzcd}
\Den(\hH) \arrow[r,dotted,"W"]\arrow{rd}{e} &  \RR\otimes R(V) \arrow{d}{\phi_p} \\
&   \ilim{}{} \RR \eE_\beta 
\end{tikzcd}
\end{equation}
We write $\rho\mapsto W_\rho$ for this assignment. It turns out that $W_\rho$, an element of the representation {\color{black}ring}, is in fact the (discrete) Wigner function of $\rho$.      
Wigner functions are introduced to the quantum computation literature in the work of Gross \cite{Gro06}.

Let us introduce the Wigner function of a density matrix. First we define the point operators
$$
A_v = |V|^{-1/2} \sum_{u\in V} b_u(v) \eta(u)
$$
where $b_u$ is the representation $V\to U(1)$ defined by $b_u(v)=\omega^{\bi(v,u)}$ ($\bi$ is the symplectic form in \ref{eq:bi}). 

\Def{{\rm The \textit{Wigner function} of $\rho$ is the function $W_\rho:V\to \RR$ defined by the equation
$$
W_\rho(v) = |V|^{-1/2}\Tr(\rho A_v).
$$
}}
We construct an element of the representation {\color{black}ring} 
$$
W_\rho = \sum_{v\in V} W_\rho(v) b_v\; \in \RR\otimes R(V)
$$
by regarding the values $W_\rho(v)$ as coefficients of the representations $b_v$.
The set $\set{A_v|\;v\in V}$ is an orthonormal basis for the $p^n\times p^n$ matrices over $\CC$ with respect to the inner product $(A,B)=|V|^{-1/2}\Tr(A^\dagger B)$, where $(-)^\dagger$ stands for the conjugate transpose. Therefore we can write
$$
\rho = \sum_{v\in V} W_\rho(v) A_v.
$$
Note that the set of Pauli observables $\set{\eta(v)|\;v\in V}$ is also an orthonormal basis with respect to the same inner product.
 
The operator inner product introduced above is related  to the inner product of characters.
The set of irreducible $\alpha$-twisted representations of a finite group $G$ constitutes an orthonormal basis of $\Cl_\alpha(G)$ with respect to the inner product
$$
(\phi,\psi) = \frac{1}{|G|} \sum_{g\in G} \phi(g)\bar\psi(g). 
$$
Let $\chi_s\in R_\beta(I)$ denote the twisted representation corresponding to $s\in \eE_\beta(I)$ under the isomorphism of Proposition \ref{pro:rep-ring}. The projector onto the common eigenspace of $\set{\eta(u)|\;u\in I}$ corresponding to the eigenvalues specified by $s$ can be written as
$$
P_s = \frac{1}{|I|} \sum_{u\in I} \bar\chi_s(u) \eta(u).
$$
Using this we can show the following.

\Lem{\label{lem:Wigner-key} 
We have
$$
(e_\rho)_I(s)= \Tr(\rho P_s) = \sum_{t\in V} (b_t|_I,\chi_s) W_\rho(t). 
$$
}
\Proof{This follows from the formula for $P_s$ and the calculation
$$\Tr(\rho\, \eta(u)) = \sum_{t\in V} W_\rho(t) b_t(u).$$}

The main result of this section is the identification of the empirical model $e_\rho$   as an element of the representation {\color{black}ring}.  
We will write $W_\rho|_I$ for the restriction of $\phi_p(W_\rho)$ to an isotropic subspace $I$.  

\Thm{\label{thm:Wigner}
Sending a density matrix $\rho$ to the element $W_\rho$ in $\RR\otimes R(V)$ defines a map
\begin{equation}\label{eq:mapWigner}
W: \Den(\hH) \to \RR\otimes R(V)
\end{equation}
that makes the diagram \ref{diagram:Wigner} commute.
}
\Proof{ The interpretation of $W_\rho|_I$ depends on the prime $p$ since the map $\phi_p$ does as well.
\begin{itemize}
\item $p>2$: In this case $\phi_p$ is induced by the restriction maps $R(V) \to R(I)$. Lemma \ref{lem:Wigner-key} implies that
$$
(e_\rho)_I(s)= (W_\rho|_I,\chi_s)
$$ 
which is what we want to prove.
\vspace{2mm}

\item $p=2$: The interpretation in this case is through the character map. The restriction is induced by the restriction maps $\RR^V \to \RR^I$. The representations $b_t$ are regarded as functions $\RR^V$. Again Lemma \ref{lem:Wigner-key} finishes the proof.
\end{itemize}
}

\Rem{\label{rem:quaiprob} {\rm As an immediate corollary of this theorem and the fact that a state is uniquely determined by its Wigner function we observe that the map $W$ is injective. By extending the coefficients to $\RR$ in the domain of \ref{eq:mapWigner}, the map $W$ becomes an isomorphism of $\RR$-vector spaces. Commutativity of diagram \ref{diagram:Wigner} implies that the map $e$ is  injective. A subtle point is that $W_\rho$ is not necessarily a probability distribution as it may assume negative values. In quantum mechanics the negativity of the Wigner function indicates a divergence from the classical behavior and this is related to contextuality as we will see next.
}}
 
\subsection{Contextuality} \label{sec:distinction}
As we emphasized in Remark \ref{rem:quaiprob} the coefficients $W_\rho(v)$ can take negative values. We write $W_\rho \geq 0$ when  $W_\rho(v) \geq 0$ for all $v\in V$, and say that the Wigner function is non-negative.
Recall that a state $\rho$ is contextual if $e_\rho$ is not in the image of the map {\color{black}(see \ref{diag:DenLim})}
$$
\theta: D \eE_\beta(\iI) \to \ilim{}{} D\eE_\beta.
$$
We will look at the odd prime and $p=2$ cases separately.

\begin{itemize}
\item $p>2$: The odd prime case is nicer since $\beta$-twisted representations coincide with ordinary representations. We can consider $D\eE_\beta(V)$ as a subset of $\RR\otimes R(V)$, and $\theta$ is the restriction of $\phi_p$ to this subset.
The Wigner function, regarded as an element of $\RR\otimes  R(V)$, can be used as a distribution when $W_\rho\geq 0$. Thus it lies in $D \eE_\beta(V)$ and furthermore satisfies $\theta(W_\rho) = e_\rho$ by Theorem \ref{thm:Wigner}. Conversely, if $\rho$ is non-contextual, i.e. there exits $d$ with $\theta(d)=e_\rho$, that means $W_\rho\geq 0$ since $d=W_\rho$. Thus we obtain a proof of the  following result, which is first proved in \cite{Howard14} for $n=1$ and generalized to all $n$ in \cite{Delfosse17}.

\Cor{\label{cor:WigPos}Assume that $p>2$. Then $W_\rho\geq 0$ if and only if $\rho$ is non-contextual.}

\item $p=2$: This case is much trickier and the Wigner function does not determine contextuality as in $p>2$. There are partial results in the physics literature, for example see \cite{DBGR15,RBDOB17}. For some $\iI$ the set $\eE_\beta(\iI)$ turns out to be empty, thus for such a  cover of contexts any state $\rho$ is strongly contextual. For example, this happens if $\iI$ contains one of the Mermin square $\iI_\square$ or Mermin star $\iI_\star$ cover of contexts (see \S \ref{sec:examples}).   
\end{itemize}


\section{Classifying space for contextuality}\label{sec:classifying}

In \cite{Coho} a topological approach is introduced to study contextuality in quantum mechanics. The essential idea is that contexts can be regarded as geometric simplices, and glued to each other to form a topological space. {\color{black} State-independent} contextuality is shown to be detected by the cohomology class $[\beta]$ that lives in the second cohomology group of a certain chain complex. In this paper we introduce a space, called the classifying space for contextuality, which realizes this chain complex. The name ``classifying space" stems from the fact that the resulting space classifies principal bundles whose transition functions, whenever simultaneously defined, are given by contexts. 
 
In this section we give the construction of the classifying space $\Bcx\iI$ for a given cover $\iI$ of contexts  and prove its basic properties. 
Homotopy-theoretic properties of this space are independently studied in \cite{O16}. We recall these properties and show how to use them in contextuality. A crucial feature of $\Bcx\iI$ is that the higher homotopy groups are non-trivial in contrast to the usual classifying space of a finite group. 
In the next section we will interpret the results of the previous section (\S \ref{sec:Wigner}) on Wigner functions in terms of the (twisted) topological $K$-theory of $\Bcx \iI$.

\subsection{Construction of the space}
Let $\iI\subset \iI(V)$ be a cover of contexts. We will construct a space out of this cover. For this the language of simplicial sets, a generalization of simplicial complexes, is most suitable.   A simplicial set consists of a sequence of sets $X_0,X_1,\cdots$ indexed by natural numbers together with face and degeneracy maps, see \cite{Friedman12}. Each set $X_n$ is referred to as the set of $n$-simplices. Face maps reduce the dimension $d_i:X_n\to X_{n-1}$, whereas the degeneracy maps increase the dimension $s_j:X_n\to X_{n+1}$. This structure is enough to obtain a topological space, and the procedure is called the geometric realization
$$
|X_\dt| = \coprod_{n\geq 0} X_n \times \Delta^n /\sim
$$
where the identification $\sim$ is generated by the face and the degeneracy maps. In effect, each combinatorial $n$-simplex is replaced by the topological simplex 
$$
\Delta^n = \set{(t_0,\cdots,t_n)\in \RR^{n+1}|\; \sum_{i=0}^n t_i =1,\;\;t_i\geq 0}
$$
and glued to each other using the face and the degeneracy maps. 

The standard example of a simplicial set is the nerve of a category. A (small) category comes with a set of objects and  a  set of morphisms. Let $\catC$ be a category. The nerve of $\catC$ is the simplicial set whose set of $n$-simplices consists of composable arrows of the form
$$
A_0 \to \cdots \to A_n.
$$
We denote the nerve by $N(\catC)$.
Face maps correspond to composing two adjacent morphisms, and degeneracy maps are given by inserting an identity morphism.

\Def{{\rm The classifying space for contextuality, denoted by $\Bcx(\iI)$, is the geometric realization  of the simplicial set $\Bcx(\iI)_\dt$  whose set of $n$-simplices is given by
$$
\Bcx (\iI)_n =\set{(v_1,v_2,\cdots,v_n)|\;  v_i\in I,\; 1\leq i\leq n, \text{ for some }I\in \iI}
$$ 
with the simplicial structure maps
$$
d_i(v_1,\cdots,v_n)= \left\lbrace
\begin{array}{ll}
(v_2,\cdots,v_n) & i=0 \\
(v_1,\cdots,v_i+v_{i+1},\cdots, v_n) & 0<i<n \\
(v_1,\cdots,v_{n-1}) & i=n
\end{array}
\right.
$$
and $s_i(v_1,\cdots,v_n)=(v_1,\cdots,v_{i-1},0,v_{i},\cdots,v_n)$ for $0\leq i \leq n$. 
}}

\Rem{\label{rem:Bcom}\rm{ This construction is analogous to a construction introduced in \cite{ACT12}. For a group $G$ one can construct a classifying space $B_\com (G)$ for commutativity. When $G$ is abelian the construction recovers the usual classifying space $BG$. A similar definition works for an arbitrary collection $\fF$ of subgroups of $G$ which is closed under taking intersections. We can define a space from the set of $n$-tuple of elements which belong to a subgroup in $\fF$. Taking $\fF$ to be the collection of all abelian subgroups of $G$ gives $B_\com(G)$. Homotopy-theoretic properties of these spaces are studied in \cite{O14,O15,O16}.

}}

Note that the construction is a variation of the standard construction of a classifying space for a group. The set of $n$-simplices of $\Bcx(\iI)$ are contained in the set of $n$-simplices, which is given by the $n$-fold direct product $V^n$, of the classifying space $BV$. The simplicial structure of $BV$ is defined similarly. We denote the inclusion map by 
$$
\iota: \Bcx(\iI) \to BV.
$$

\Rem{\label{rem:Princx} {\rm

The name ``classifying space" refers to the fact that $BG$ classifies principal $G$-bundles \cite{Steenrod}. More precisely, given a CW-complex $X$ the set $[X,BG]$ of homotopy classes of maps  is in one-to-one correspondence with the set of isomorphism classes of principal $G$-bundles over $X$. Similarly $B_\com G$ is a classifying space for certain types of principal bundles. These bundles have the property that the transition functions commute whenever simultaneously defined \cite{AG15}. A similar conclusion can be made for  $\Bcx \iI$. This space classifies principal $V$-bundles whose transition functions, whenever simultaneously defined, specifies a context. We denote the equivalence classes of such bundles by $\Princx^V(X)$.

}}

\subsection{Chain complex}
The chain complex  $C_*(V)$ of the classifying space $BV$ can be used to compute the group homology of $V$, see for instance \cite{Brown12}.
The chain complex of $\Bcx(\iI)$ 
is a subcomplex of   $C_*(V)$, and the complex used in \cite{Coho} consists of the part where the dimension is $\leq 3$. The full complex is infinite-dimensional in general.

For contextuality we will use  $\ZZ/p$ coefficients at all times, so we omit the coefficients and simply write $C_*(\iI)$ for the chain complex of $\Bcx\iI$.  Then $C_n(\iI)$ is the $\ZZ/p$-vector space generated by the set $\Bcx(\iI)_n$. The boundary map is the alternating sum of the face maps: 
$$
\partial[v_1|\cdots|v_n] = [v_2|\cdots|v_n]+\left(  \sum_{i=1}^{n-1} [v_1|\cdots|v_i+v_{i+1}|\cdots|v_n] \right) + (-1)^n[v_1|\cdots|v_{n-1}]. 
$$  
This complex, for $n=0,1,2,3$, is introduced in \cite{Coho}.
There is a corresponding cochain complex denoted by $C^*(\iI)$. It can be obtained from $C_*(\iI)$ by applying $\Hom(-,\ZZ_p)$ to the chain complex. More explicitly, the $n$-cochains can be thought of as functions $\alpha:\Bcx(\iI)_n \to \ZZ/p$. The coboundary is defined by $d\alpha(x)=\alpha(\partial(x))$ where $x\in  C_{n+1}(\iI)$.

We write $H_*(-)$ for the homology of a space with mod-$p$ coefficients. The homology group $H_*(\Bcx\iI)$ coincides with the homology of the chain complex $C_*(\iI)$. Similarly we write $H^*(-)$ for the cohomology group with mod-$p$ coefficients, and $H^*(\Bcx\iI)$ coincides with $H^*(C^*(\iI))$.
 
Let us consider the cohomology group $H^1(\Bcx\iI)$. By definition $C^1(\iI)=C^1(V)$. Therefore it follows that 
\begin{equation}\label{eq:iotaH1}
\iota^*:H^1(BV) \to H^1(\Bcx\iI)
\end{equation}
is injective. Choosing a symplectic basis $\set{z_i,x_i}_{i=1}^n$ for $V$ the  $\ZZ/p$-vector space it can be shown that $H^1(BV)$ is spanned by the dual basis denoted by $z_i^*$ and $x_i^*$ \cite{Brown12}. Here $z_i^*:V\to \ZZ/p$ is the homomorphism which sends $z_i$ to $1$ and the rest of the basis elements to $0$  ($x_i^*$ is defined analogously). We will see that $\iota^*$ in degree $1$ is an isomorphism for $p>2$, but when $p=2$ the image has an extra $\ZZ/2$ factor.

Recall from Remark \ref{rem:betaI} that the extension cocycle $\beta$ when restricted to a context $I$ can be regarded as a function $\beta|_I:I\times I \to \ZZ/p$. Since $\Bcx(\iI)_2$ consists of the union of $I\times I$ where $I\in \iI$ we can regard $\beta$ as a $2$-cochain on $\Bcx(\iI)$:
$$
\beta: \Bcx(\iI)_2 \to \ZZ/p.
$$
The set $\eE_\beta(\iI)$ can be described as the preimage of $\beta$ under the coboundary map $d:C^1(\iI)\to C^2(\iI)$. 
 
\Pro{\cite{Coho} If $[\beta]\neq  0$ in $H^2(\Bcx\iI)$ then any state $\rho$ is strongly contextual.}
\Proof{ Unraveling the definition, $[\beta]\neq 0$ means that $\eE_\beta(\iI)$ is empty. Then Corollary \ref{cor:contextEbeta} implies that $\rho$ is strongly contextual. 
}

\subsection{The distinction between $p=2$ and $p>2$} In contextuality the behavior of the odd and even prime cases is very different. This distinction, first appeared in \S \ref{sec:distinction}, can now be given a cohomological interpretation.

We start our analysis with the odd prime case as it is relatively easy. The group $Z_\mu$ is isomorphic to $\ZZ/p$. Thus the coefficient group in $H^2(BV)$ defining the extension $P_n$ and the coefficient group used in $H^2(\Bcx\iI)$ is the same.   Proposition \ref{pro:beta-canonical}, which says that the canonical class $\beta \in C^2(\iI)$ is the zero function, {\color{black} implies that the class $[\beta]$ in $H^2(\Bcx\iI)$ vanishes. Therefore $[\beta]$ cannot detect strong contextuality when $p>2$.}



Now we turn to $p=2$ case. Recall that $Z_\mu$ is isomorphic to $\ZZ/4$. The change of coefficient between the cohomology group $H^2(BV,Z_\mu)$, which contains the extension class, and the cohomology group $H^2(\Bcx\iI)$, which detects strong contextuality, requires a more careful consideration. In the physics literature it is well-known that there are strongly contextual situations independent of the chosen state {\color{black}(state-independent contextuality)}. The simplest such case is  Mermin's square as we will discuss in \S \ref{sec:examples}.  

It will be useful to describe $[\beta]$ more explicitly. We will give a cup product decomposition in terms of $1$-dimensional classes. We begin by describing the extra factor in the image of $\iota^*$ in \ref{eq:iotaH1}. Since $\iota^*$ is injective any group homomorphism $V\to \ZZ/2$ gives a non-zero element in $H^1(\Bcx\iI)$. There is also an extra element which does not come from a homomorphism. Let $\qu:V\to \ZZ/2$ denote the function $\qu(v)=v_x\cdot v_z$. ({{\color{black} Composing $\qu$ with  the  inclusion $\ZZ/2=\set{0,1} \subset \ZZ/4$ gives the function used in \ref{eq:beta-canonical}.}) By definition this function satisfies $d\qu =\bi$ as cochains in $ C^2(V)$. Thus $d\qu=0$ when restricted to $C^2(\iI)$. But this is not true in general: $\qu(v)+\qu(v')\not= \qu(v+v')$ if $\bi(v,v')\not=0$. Therefore there is a corresponding cohomology class in $H^1(\Bcx\iI)$, which will   be denoted by $\qu$.

\Pro{\label{pro:cup}As an element of $H^2(\Bcx\iI)$ the class of $\beta$ can be expressed as
$$
[\beta] = \qu^2 + \sum_{i=1}^n x^*_i\cup z^*_i.
$$
}
\Proof{ We will use the section  $\eta(v)=i^{\bar\qu(v)}Z(v_z)X(v_x)$ that appears in the definition \ref{eq:beta-canonical} of the canonical choice for $\beta$. {\color{black} We can think of $\bar\qu$ as the function}
$V\to \ZZ/4$ obtained by regarding $\qu(v)\in \set{0,1}$ as taking values  in $\ZZ/4$. The corresponding cocycle is   $\beta= \beta_0+d\bar\qu$ with $\beta_0(v)=2v_x\cdot v_z$.  We will compute the coboundary $d\bar\qu$. To be able to distinguish mod $2$ and  mod $4$ additions we regard $\ZZ/4$ as the extension $\ZZ/2\times_\tau\ZZ/2$ defined using the cocycle $\tau(t,t')=tt'$. We write the multiplication as $(t_1,t_2)(t_1',t_2')=(t_1+t_1'+t_2t_2',t_2+t_2')$. Then the coboundary  gives
\begin{align*}
d\bar\qu(v,v') &= (0,\qu(v))\,(0,\qu(v'))\,(0,\qu(v+v'))^{-1}\\
&= (\qu(v)\qu(v'), \qu(v)+\qu(v'))\,(\qu(v+v'),\qu(v+v'))\\
 &= (\qu(v)\qu(v'), \qu(v)+\qu(v')+\qu(v+v'))\\
  &= (\qu(v)\qu(v'), 0)
\end{align*}
where we used $(0,t)^{-1} = (t,t)$ and $d\qu = \bi \mod 2$. In $\ZZ/4$  this equation is expressed as $d\bar\qu = 2\bar\qu^2$. Thus we obtain the desired cup product decomposition $\beta = 2\bar\qu^2 + 2\sum_{i=1}^n x_i^*\cup z_i^*$ after identifying $2(\ZZ/4)\cong \ZZ/2$, which amounts to dividing by $2$. 
}

\subsection{Homotopy theory} The space $\Bcx\iI$ is studied in \cite{O16}. Formally, there is a structural similarity to the space $B_\com G$ (see \ref{rem:Bcom}), which has been studied in \cite{O14,O15}, and most of the  tools developed there also apply to $\Bcx\iI$. We sketch the relevant homotopical properties whose proofs can be found in these papers.

By construction $\Bcx(\iI)$ can be expressed as the union of classifying spaces
$$
\Bcx(\iI) =  \bigcup_{I\in \iI} BI.
$$
This can be seen from the description of $n$-simplices. We can also think of the union as a colimit in the category of spaces. 
To study homotopy-theoretic properties   the union is usually replaced by a homotopy colimit \cite{Dugger}---a gluing process that has nice homotopical properties. We can define a functor $B:\iI\to \catTop$ that sends $I$ to the classifying space $BI$. There is a natural map
$$
\hocolim{\iI} B \to \Bcx(\iI)
$$
which turns out to be a homotopy equivalence, see \cite{O14}. 
This approach also helps us to compute cohomology of the space by using a spectral sequence. Cohomology (or even generalized cohomology) of a homotopy colimit can be computed using a spectral sequence.

\Ex{\label{ex:n=1} {\rm
Let us explain the simplest case of $n=1$. Let $X\vee Y$ denote the wedge sum of the spaces $X$ and $Y$ at a specified basepoint. The homotopy colimit decomposition for $n=1$ is simply given by
$$
\Bcx \iI(V) \simeq \bigvee^{r_p } B\ZZ/p.
$$
where $r_p=(p^{2n}-1)/(p-1)$ is the number of maximal isotropic subspaces. {\color{black} This case falls into the   class of transitionally commutative groups studied in \cite{ACT12}.}
}}

Another benefit that we gain by switching to homotopy colimits is a formula for the fundamental group
\begin{equation}\label{eq:presentation}
\pi_1(\Bcx\iI) = \Span{e_v,\;v\in\Sigma|\; e_v e_{v'}=e_{v+v'}\;\text{ if } \set{v,v'}\subset I \text{ for some } I\in\iI }.
\end{equation}
We will denote the fundamental group by $\pi(\iI)$, or just by $\pi$ if the context is understood. Note that each isotropic subspace $I\in \iI$ can be regarded as a subgroup of $\pi$ by sending an element $u$ to the corresponding generator $e_u$. Indeed, there is a set map $\Sigma \to \pi$. Moreover, this map induces a map of spaces
$$
\varphi: \Bcx\iI \to B\pi
$$
that is defined by sending an $n$-simplex $(v_1,\cdots,v_n)$ to the tuple $(e_{v_1},\cdots,e_{v_n})$. It is an interesting fact that $\varphi$ is not a homotopy equivalence in general. In other words, the higher homotopy groups $\pi_i(\Bcx\iI)$, $i\geq 2$, can be non-trivial in contrast to $B\pi$ whose homotopy groups are zero except $\pi_1(B\pi)=\pi$.

From the homotopical point of view all the information about the higher homotopy groups is hidden in the homotopy fiber of $\varphi$. It turns out that the homotopy fiber has a very concrete description in terms of a combinatorial object called the coset poset. 

\subsection{Coset poset}\label{coset-poset}
We introduce this space for an arbitrary group $G$. 
\Def{{\rm
For a collection of subgroups $\fF$ of $G$ we define the poset
$$
C_G(\fF) =  \set{gA|\; A\in \fF,\;\; g\in G }
$$
ordered by the inclusion relation.
}}
 
We usually identify the poset with the associated space given by   the geometric realization of its nerve. Our convention for the morphisms of the category associated to a poset $\pP$ is that we write $A\to B$ if $A\leq B$. The associated space has $n$-simplices given by composable morphisms $A_1\to\cdots\to A_n$.

Given this construction and the fundamental group $\pi(\iI)$ of $\Bcx(\iI)$ we can describe the homotopy fiber of $\varphi$. Each subspace in the context $\iI$ can be regarded as subgroup of $\pi$, thus we can construct the coset poset $C_{\pi} (\iI)$. 
The main result of \cite{O16} identifies the homotopy type of $C_{\pi} (\iI)$ when $\iI=\iI(V)$. In the simplest case $V=(\ZZ/p)^2$ the coset poset $C_\pi(\iI)$ turns out to be contractible. As a consequence, in this case $\varphi$ is a homotopy equivalence. For $n\geq 2$ things get interesting:

\Thm{\cite{O16} \label{thm:homotopy-groups}
Consider $\iI=\iI(V)$ where $V=(\ZZ/p)^{2n}$ and $n\geq 2$.
The fundamental group $\pi_1(\Bcx\iI)$ is given by the central extension associated to the cocycle $\bi$
$$
\pi  \cong V\times_\bi \ZZ/p 
$$ 
 and the higher homotopy groups are  given by
$$
\pi_i (\Bcx\iI) \cong  
\pi_i(\bigvee^{d(p,n)} S^n)  
$$
for all $i\geq 2$, where  the number of spheres is given by the formula
$$
d(p,n)=(-1)^{r+1}+p^{2r+1+r^2}+ \sum_{j=1}^r (-1)^{j} p^{2r+1-j+(r-j)^2} \left( \prod_{t=0}^{j-1} \frac{p^{2r-t} - p^{t}}{p^j-p^t} \right)
$$
}

\Rem{{\rm
We will see some applications of this result. However, we do not know of an interpretation for the numbers $d(p,n)$.  
This number gives information about  the global structure of the classifying space for contextuality. Interpreted in terms of principal bundles we have
$$
\Princx^V(S^n) = \ZZ^{d(p,n)}
$$  
where we used the notation of Remark \ref{rem:Princx}. On the other hand, The set $\Prin^V(S^n)$ of equivalence classes of ordinary principal bundles over the $n$-sphere, $n\geq 2$, has a single element given by the trivial bundle. 
}}

\Rem{\label{rem:pi1}{\rm 
A better way to rephrase Theorem \ref{thm:homotopy-groups} is to say that for $n\geq 2$ there is a fibration sequence
\begin{equation}\label{eq:fibration}
\bigvee^{d(p,n)} S^n \to \Bcx(\iI) \stackrel{\varphi}{\longrightarrow} B\pi.
\end{equation}
Moreover, the group $\pi$ turns out to be
$$
V\times_\bi \ZZ/p  \cong \left\lbrace
\begin{array}{cc}
V\times \ZZ/2 & p=2\\
P_n & p>2
\end{array}\right.
$$
as a consequence of the presentation given in \ref{eq:presentation}. 
}}

\subsection{Cohomology revisited} Having the homotopical description of $\Bcx(\iI)$ we can improve our statements about cohomology. The first application is to determine the extra factor in $H^1(\Bcx\iI)$ when $p=2$.

\Cor{Assume $n\geq 2$. There is an isomorphism 
$$
H^1(\Bcx\iI)\cong \left\lbrace 
\begin{array}{cc}
(V\times \ZZ/2)^* & p=2 \\
V^* & p>2
\end{array}
\right.
$$
where $(-)^*$ stands for the dual vector space.
} 
\Proof{
This follows from the description of the fundamental group $\pi$, and the standard fact that for a connected space $X$ the first homology group is the abelianization of the fundamental group \cite{Hat02}.
}

The Serre spectral sequence \cite{Hat02} of the fibration \ref{eq:fibration} gives us an isomorphism
$$
\varphi^*:H^i(B\pi)\to  H^i(\Bcx\iI) 
$$
in degrees $i\leq n-1$ and an injection in degree $i=n$. When $p=2$ we have $\pi=V\times \ZZ/2$ as observed in Remark \ref{rem:pi1}, so its cohomology is easy to describe. Fixing the basis $\set{x_i,z_i|\;1\leq i\leq n}\cup \set{x_0}$ for $V\times \ZZ/2$ the cohomology ring is given by the polynomial ring
$$
H^*(B\pi) = \FF_2[x_0^*,x_1^*,\cdots,x_n^*,z_1^*,\cdots,z_n^*].
$$ 

Let us define a homomorphism 
$$
q: \pi\to V\times \ZZ/2 \;\;\;\; (v,t)\mapsto (v,t+\qu(v))
$$
which turns out to be  an isomorphism of groups. Now consider the composite map of spaces
$$
\hat q:\Bcx\iI \stackrel{\varphi}{\longrightarrow} B\pi \stackrel{B q}{\longrightarrow} B(V\times \ZZ/2) 
$$
As a consequence of the cup product decomposition given in Proposition \ref{pro:cup} we obtain:

\Cor{\label{cor:Q-beta} The cohomology class $[\beta]\in H^2(\Bcx\iI)$ is the image of
$$
Q = x_0^* + \sum_{i=1}^{n} x_i^*\cup z_i^* \;\; \in H^2(B(V\times \ZZ/2))
$$  
 under the map  $H^*(B(V\times \ZZ/2))\to H^*(\Bcx\iI)$ induced by $\hat q$.
} 

It is well-known that $Q$ is the extension class of $P_n$, the extraspecial $2$-group of complex type, when regarded as the central extension
$$
0 \to \ZZ/2 \to P_n \to V\times \ZZ/2 \to 0
$$
where $\ZZ/2$ (isomorphic to {\color{black}$\Span{-I_{2^n}}$}) is the commutator subgroup \cite{Q71a}. We can think of $Q$ as a quadratic form on $V\times \ZZ/2$. Under the homomorphism $q$ an isotropic subspace of $V$ is mapped to a singular subspace of $V\times \ZZ/2$. A \textit{singular subspace} is an isotropic subspace $S$ with respect to the bilinear form associated to $Q$ such that $Q|_S=0$.

\section{Contextuality and twisted $K$-theory} \label{sec:K-theory}
 
The goal of this section is to 
 compute the $\beta$-twisted   $K$-theory of $\Bcx \iI$. The idea follows the computation of the topological $K$-theory of $B_\com G$ given in \cite{O14}. We refer to this paper for the details of the constructions involved. Another reference that is useful for the homotopical properties of this space is \cite{O16}.
The computation will bridge a connection to the representation-theoretic description of the Wigner function given in \S \ref{sec:Wigner}. This will allow us to interpret the Wigner function as a class $[W_\rho]$  in the $\beta$-twisted $K$-theory of $\Bcx \iI(V)$ after extending the coefficients to $\RR$. 
 
\Rem{\rm Another form of $K$-theory, namely
operator $K$-theory, has been used in the study of contextuality by de Silva and Barbosa \cite{Silva18}. Therein  the authors relate operator $K$-theory to topological $K$-theory via the Gelfand spectrum. However, there seems  to be no direct connection to the way $K$-theory appears in the present work. }

\subsection{Sketch of the computation}
Before embarking on the details, let us first   summarize the idea of the proof. 
As an element of the inverse limit of $\RR\otimes R_\beta$ the Wigner function is a sum 
$$
W_\rho = \sum_{v\in V} W_\rho(v)\, b_v
$$
of $1$-dimensional irreducible representations of the group $V$. Such a sum is usually referred to as a ``virtual" sum when the coefficients  assume negative values. In this section we will explain a chain of ideas, due to Atiyah and Segal, that will allow us to pass from twisted representations to twisted $K$-theory. This is given by the ``completion" of the representation {\color{black}ring} in the sense that Cauchy sequences are convergent in the resulting completed module. At the level of spaces the completion process corresponds to a construction known as the Borel construction. Let $G$ be a group and let $EG\to BG$ denote the universal principal $G$-bundle \cite[Chapter II]{adem2013cohomology}. For a space $X$ with an action of  $G$ the Borel construction is defined to be the space
$$
EG\times_G X = (EG\times X)/G
$$ 
obtained as the quotient  under the diagonal action. 
The Atiyah--Segal completion theorem \cite{AS69} says that the 
topological $K$-theory group $K(EG\times_G X)$ is the completion of the equivariant $K$-theory group $K_G(X)$. The equivariant $K$-theory group is a generalization of the representation {\color{black}ring}. When $X$ is a point with the trivial $G$-action then $K_G(X)=R(G)$.  All of this is also true in the twisted setting \cite{Lah12}. We are only interested in twisting by the cocycle $\beta$.

 In our case we take $X$ to be the coset poset $\cC_\pi \iI$ with the action of $\pi$ given by left multiplication on the cosets. The Borel construction gives us $\Bcx\iI$, as shown in  \cite{O16}. Therefore by the Atiyah--Segal completion theorem  the twisted $K$-theory of $\Bcx\iI$ is the completion of the twisted equivariant  $K$-theory of the coset poset.  There  are two steps of the computation
\begin{itemize}
\item After extending the coefficients to $\RR$ we show that $K_\pi^\beta(\cC_\pi\iI)$ is equal to  the inverse limit of $\RR\otimes R_\beta$. Thus $W_\rho$ belongs to the $\beta$-twisted equivariant $K$-theory group.
\vspace{5mm}

\item Completing $K_\pi^\beta(\cC_\pi\iI)$  we obtain $K^\beta(\Bcx\iI)$ and after extending coefficients to $\RR$ we  conclude that   the Wigner function can be regarded as a class $[W_\rho]$ in the $\beta$-twisted $K$-theory group.
 \end{itemize}

\subsection{Twisted  equivariant  $K$-theory} We can twist the equivariant $K$-theory groups by the cohomology class $[\beta] \in H^2(\Bcx\iI(V))$. Recall that $[\beta]=0$ when $p$ is odd. Thus in this case twisting disappears. For $p=2$ we observed in Corollary \ref{cor:Q-beta} that $[\beta]$ comes from a class in $H^2(B\pi,\ZZ/2)$. We will treat the two cases in a uniform way. Therefore we think of $[\beta]$ as living in $H^2(B\pi,\ZZ/p)$ given by the class $Q$ when $p=2$, and understood to be zero when $p>2$.  We assume $n\geq 2$ so that $\pi$ is a finite group (Remark \ref{rem:pi1}). The case $n=1$ can be computed directly, and we defer this case to the proof of Theorem \ref{thm:K-theory-compute}.

As observed in \cite{dwyer2008twisted} twisted equivariant $K$-group can be obtained from the untwisted version. 
To see this let us consider the group extension
$$
0 \to \ZZ/p \to \pi_\beta \to \pi \to 0
$$
corresponding to  $[\beta]$. Note that the coset poset $C_\pi \iI$ can be regarded as a $\pi_\beta$-space via the  quotient homomorphism $\pi_\beta\to \pi$.
Irreducible representions of $\ZZ/p$ consist of the homomorphisms $\omega^j:\ZZ/p\to U(1)$ sending the additive generator $1$ to the element $\omega^j$ where $0\leq j \leq p-1$.   Then we have
\begin{equation}\label{eq:directsum}
K^{*}_{\pi_\beta} (C_\pi \iI) =  \bigoplus_{j=0}^{p-1}  K^{*}_{\pi_\beta} (C_\pi \iI)(\omega^j)
\end{equation}
where the factors correspond to equivariant bundles on which $\ZZ/p$ acts by the specified representation.
 We can identify the twisted equivariant groups $K^{\beta+*}_\pi(C_\pi\iI)$ with the factor corresponding to the $j=1$ representation:
$$
K^{\beta+*}_\pi (C_\pi\iI) = K^{*}_{\pi_\beta} (C_\pi \iI)(\omega). 
$$
Note that this representation corresponds to  the usual embedding   $\ZZ/p\subset U(1)$. 

Equivariant $K$-theory of the coset poset can be computed using the methods of \cite{O14}.  The main tool is a spectral sequence that can be applied to homotopy colimit decompositions. As shown in \cite{O16}
the coset poset is the homotopy colimit of the functor $\pi/-: \iI\to \catTop$ which sends $I$ to the coset $\pi/I$ regarded as a discrete space. Furthermore, as a consequence of \ref{eq:directsum} the spectral sequence decomposes and we can use it to compute the twisted equivariant $K$-groups. Therefore   the results for the twisted version are essentially obtained in the same way as the untwisted case.

Before discussing the spectral sequence we need a construction that extends the construction of the inverse limit of a functor. Let $F:\iI^\op\to \catAb$ be a   functor that takes values in  the category of abelian groups. One can construct a cochain complex $C^*(\iI,F)$ which is indexed over the $n$-simplices of the nerve of the poset $\iI$:
$$
C^n(\iI,F) = \bigoplus_{A_1\to\cdots \to A_n} F(A_1)
$$
and the boundary map comes from the simplicial structure of the nerve of $\iI$. The cohomology group $H^i(\iI,F)$ gives the $i$-th derived functor of the inverse limit of $F$. Note that $H^0(\iI,F)$ is canonically isomorphic to the inverse limit of $F$. 

{\color{black}
\Rem{{\rm
We can also think of the groups $H^i(\iI,F)$ as the \v{C}ech cohomology groups of the presheaf $F$ with respect to the cover $\iI$. These cohomology groups are closely related to the \v{C}ech cohomology groups considered in \cite{abramsky2011cohomology}. Therein the choice of $F$ is the functor $\RR S_d: \iI^\op \to \catVect_\RR$ sending an isotropic subspace $I$ to the  $\RR$-vector space $\RR S_d(I)$  of  
$\RR$-linear combinations of $s:I\to \ZZ/p$ in the support of $d_I$, i.e. those $s$ satisfying $d_I(s)\neq 0$. Note that when $d=e_\rho$ for some state $\rho$ the presheaf $\RR S_d$ is a subpresheaf of $\RR \eE_\beta$ as a consequence of Proposition \ref{pro:Ebeta}. Therefore there is a comparison map between the \v{C}ech cohomology groups
$$
H^i(\iI,\RR S_d ) \to H^i(\iI,\RR \eE_\beta ).
$$ 
Below we will consider $\RR\otimes R_\beta$, the functor isomorphic to   $\RR \eE_\beta$  (Proposition \ref{pro:rep-ring}).
}}
}

For homotopy colimits we can use  the (equivariant) Bousfield-Kan spectral sequence---the spectral sequence used in \cite{O14} for the untwisted case. As a consequence of the direct sum decomposition in \ref{eq:directsum} the spectral sequence decomposes:
$$
E_2^{i,j}= H^i(\iI,K_\pi^{\beta+j}(\pi/-)) \Longrightarrow K_\pi^{\beta+*}(C_\pi\iI). 
$$
The functor $K_\pi^{\beta+j}(\pi/-)$ can be identified with the $\beta$-twisted representation group functor $R_\beta:\iI^\op \to \catAb$ for $j$ even and it is  zero otherwise. This is the key connection to \S \ref{sec:Wigner}.

\Rem{\rm{We would like to remind  that $[\beta]=0$ for the odd prime case (Remark \ref{rem:untwist}). Therefore the twisting disappears in this case, and we are dealing with untwisted $K$-groups.}}

We are interested in understanding the spectral sequence after tensoring with $\RR$. The following serves to this purpose.
  
\Lem{\label{lem:torsion} The derived limit functor $H^i(\iI, R_\beta)$ is torsion for $i>0$.}   
\Proof{
This follows from \cite[Theorem 4.7]{O14} once we show $ R_\beta$ is a pre-Mackey functor. Restriction and induction of twisted characters provide the necessary structure \cite[Theorem 4.2]{dwyer2008twisted}.
}

As a consequence of this result  our spectral sequence collapses after tensoring with $\RR$. Here we are also using the fact that the spectral sequence is bounded in the $i$-direction---all the terms vanish for sufficiently large $i$.
For $i>0$   the terms $\RR\otimes E_2^{i,j}$ are zero since $ E_2^{i,j}$ are torsion as proved above. When $i=0$ and $j$ even we have $\RR\otimes E_2^{0,j} =   H^0(\iI,\RR\otimes R_\beta)$, and for $j$ odd $\RR\otimes E_2^{0,j}=0$. Recall that $H^0(\iI,\RR\otimes R_\beta)$ can be identified with the inverse limit of the functor $\RR\otimes R_\beta$.
For $\iI=\iI(V)$ we have computed this inverse limit in Proposition \ref{pro:inv-lim-rep}. Thus we obtain
\begin{equation}\label{eq:equivK}
\RR\otimes K_\pi^{\beta+i}(C_\pi \iI) \cong \left\lbrace 
\begin{array}{cc}
\RR\otimes R(V) & i=0 \\
0 & i=1. 
\end{array}
\right.
\end{equation} 

Therefore we can take the target of the function $W$ in \ref{eq:mapWigner}, that sends a state $\rho$ to the Wigner function $W_\rho$, to be the twisted equivariant $K$-theory group:
\begin{equation}\label{eq:W-equivariant}
W: \Den(\hH) \to \RR\otimes K_\pi^\beta(C_\pi \iI(V)).
\end{equation}
We can express this by saying that a state $\rho$ can be interpreted as a probabilistic combination of $\pi$-equivariant $\beta$-twisted vector bundles over $C_\pi \iI$. Our next goal is to relate this to the twisted $K$-theory of the classifying space for contextuality.

\subsection{Twisted $K$-theory}  
{\color{black}  
Let $I(\pi)$ denote the augmentation ideal, the kernel of the   homomorphism $R(\pi) \to \ZZ$ which sends a representation to its dimension.}
The twisted version \cite{Lah12} of Atiyah-Segal completion theorem gives an isomorphism
$$
K^{\beta+*}(\Bcx\iI) \cong K_\pi^{\beta+*}(C_\pi \iI)^\wedge
$$
where $(-)^\wedge$ means completion with respect to the augmentation ideal $I(\pi)$.  

Let us explain the completion process. First of all $K_\pi^{\beta+*}(C_\pi \iI)$ is a module over the representation ring $R(\pi)$. This is the case since $K_{\pi_\beta}^*(C_\pi\iI)$ is a module over the ring $R(\pi_\beta)$ via the map $C_\pi \iI\to \ast$ which contracts the coset poset to a point, and using the ring map $R(\pi)\to R(\pi_\beta)$ induced by the quotient homomorphism we see that the $\beta$-twisted $K$-group is a  module over $R(\pi)$.
Then the completion can be computed by   tensoring with the completion of $R(\pi)$:
$$
K_\pi^{\beta+*}(C_\pi \iI)^\wedge \cong K_\pi^{\beta+*}(C_\pi \iI) \otimes_{R(\pi)} R(\pi)^\wedge.
$$
Therefore it suffices to understand the completion of $R(\pi)$. Recall from Remark  \ref{rem:pi1} that $\pi$ is a $p$-group. It is well-known that for $p$-groups completion of the representation ring turns out to be essentially tensoring with the  $p$-adic numbers $\ZZ_p$:
$$
R(\pi)^\wedge \cong \ZZ\oplus(\ZZ_p\otimes I(\pi)).
$$

\Thm{\label{thm:K-theory-compute}There is an isomorphism
$$
\RR\otimes K^{\beta+i}(\Bcx\iI(V)) \cong \left\lbrace 
\begin{array}{cc}
\RR \oplus \left( \RR\otimes \ZZ_p^{|V|-1} \right) & i=0 \\
0 & i=1.
\end{array}
\right.
$$
}
\Proof{
The proof  follows the proof of the untwisted version in  \cite[Theorem 5.2]{O14}. For $n=1$ the result is a direct consequence of the decomposition in Example \ref{ex:n=1}. Assume $n\geq 2$. The order of completion and tensoring with $\RR$ matters. We first complete then tensor with $\RR$. Lemma \ref{lem:torsion} says that for $i>0$ the groups $E_2^{i,j}$ are torsion. Under the completion process, which is essentially tensoring   with $\ZZ_p$, these groups do not change. After tensoring with $\RR$ they disappear. Therefore we focus on $i=0$ part. Completion of $ H^0(\iI,R_\beta)$ gives us $\ZZ\oplus \ZZ_p^{|V|-1}$ (see Remark \ref{rem:free-abelian}). Finally tensoring with $\RR$ gives the desired result. 
} 

As a consequence of this result we can think of the  map $W$ in \ref{eq:W-equivariant} as landing inside twisted $K$-theory
$$
W: \Den(\hH) \to  \RR\otimes K^\beta(\Bcx\iI(V)).
$$
Note that the completion map 
$$
 K_\pi^{\beta}(C_\pi \iI(V)) \to  K^{\beta}(\Bcx\iI(V)) 
$$
is injective since tensoring with $\ZZ_p$ is. This means that we do not lose any information during the completion process and can identify the Wigner function $W_\rho$ as a class in the $\beta$-twisted $K$-group.

\section{Examples} \label{sec:examples}

We study two examples whose  constructions are due to Mermin \cite{Mer93}. They  are usually referred to as Mermin star and Mermin square. We use these examples to illustrate the use of $\Bcx\iI$ and give a flavor for Theorem \ref{thm:homotopy-groups}. Then we turn to {\color{black}Mermin's inequality \cite{mermin1990extreme}}---a type of Bell inequality. We use our representation-theoretic approach to generalize  the well-known Mermin inequality obtained from the Mermin star construction along the lines of \cite{Fraction2018}.

\subsection{Mermin's construction}  This example can be regarded as an illustration of Theorem \ref{thm:homotopy-groups} in a simpler situation. 
In Fig.\ref{fig:Mermin} we see an arrangement of the observables on the edges of a torus, taken from \cite{Coho}. Unlabeled edges are determined from the others: in each triangle the sum of the labels on the boundary gives $0$. 

\begin{figure}[h!]
  \includegraphics[scale=0.7]{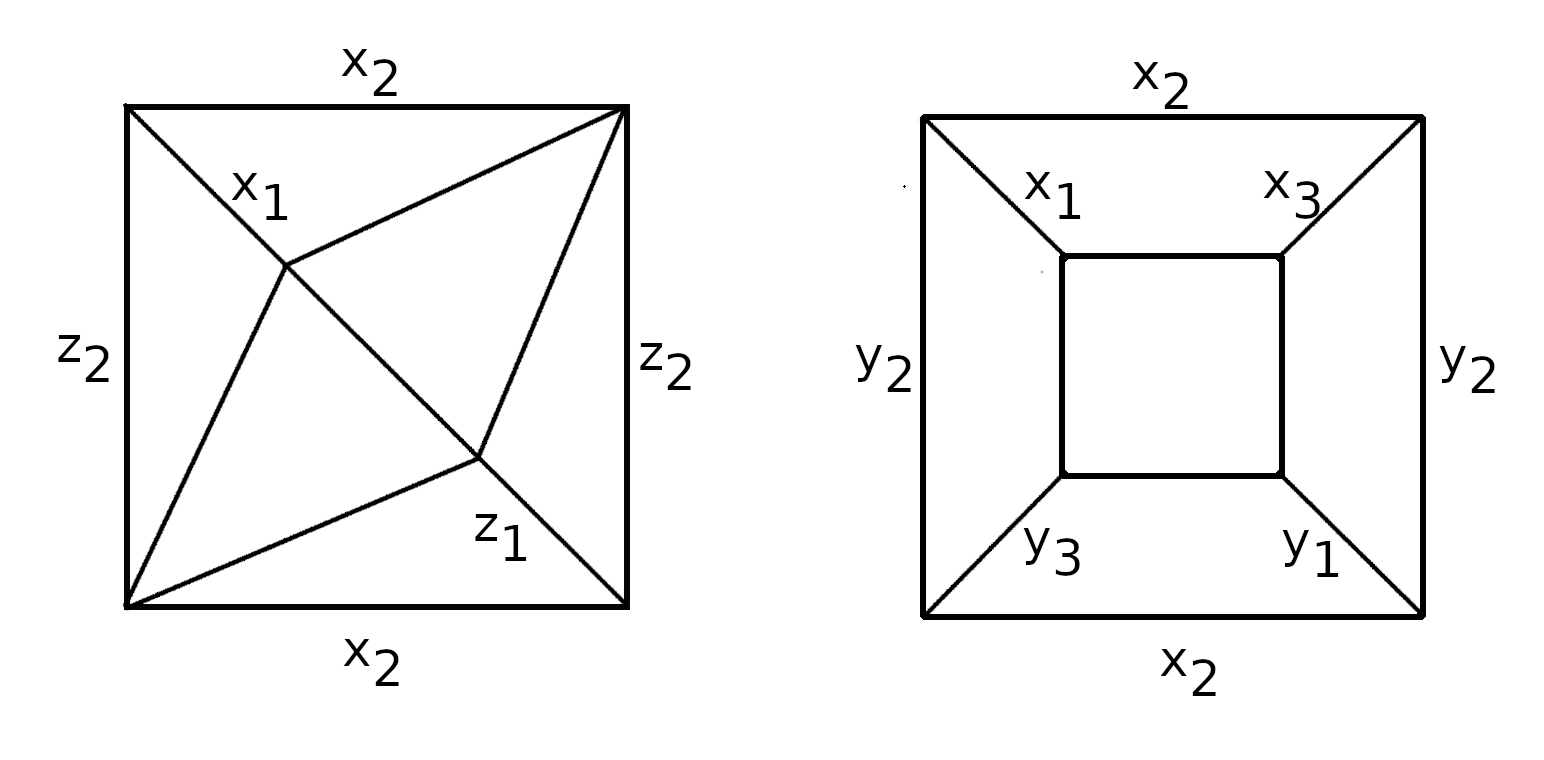}
  \caption{Mermin square and star observables as labels on the edges of a torus.}
  \label{fig:Mermin}
\end{figure}

\begin{itemize}
\item Mermin square is a $2$-qubit strongly contextual set of observables. It is constructed from the collection  
\begin{align*}
I_1 & = \Span{x_1,x_2} \\
I_2 & = \Span{x_1+x_2,z_1+z_2} \\
I_3 & = \Span{z_1,z_2}\\
I_4 & = \Span{x_1,z_2} \\
I_5 &= \Span{x_1+z_2,z_1+x_2}\\
I_6 &= \Span{z_1,x_2}
\end{align*} 
The cover $\iI_\square$ of contexts consists of $I_i$, pairwise intersections $I_i\cap I_j$, and triple intersections $I_i\cap I_j\cap I_k =\set{0}$. A convenient way of arranging these observables is illustrated in Fig.\ref{fig:Mermin}.
\vspace{5mm}

\item Mermin star consists of observables for $3$-qubits.   The contexts are given by 
\begin{align*}
I_1 & = \Span{x_1,x_2,x_3} \\
I_2 & = \Span{x_1,y_2,y_3} \\
I_3 & = \Span{y_1,x_2,y_3}\\
I_4 & = \Span{y_1,y_2,x_3} \\
I_5 &= \Span{x_1+x_2+x_3,y_1+y_2+x_3,x_1+
y_2+y_3}
\end{align*}
where we write $y_i=x_i+z_i$ to simplify the notation.
 The cover $\iI_\star$ of contexts consists of intersections of $I_i$ in the list. The topological realization also gives a torus, see Fig.\ref{fig:Mermin}.
\end{itemize}

We will describe the classifying space for contextuality for these examples. There are two parts of this description. First we calculate the fundamental group of the space using the presentation \ref{eq:presentation}. For this the configuration depicted in Fig.\ref{fig:Mermin} helps. Then we need to understand the homotopy type of the coset poset $C_\pi \iI$.

\Pro{\label{pro:merfibseq} There are fibration sequences
$$\bigvee^{15} S^2 \to \Bcx \iI_\square \to BV_2,\;\;\;\;\bigvee^{103} S^2 \to \Bcx\iI_\star \to BV_3.$$
In particular, both contexts are strongly contextual.
}
\Proof{Recall that $\varphi:\Bcx(\iI)\to B\pi$ has a homotopy fiber given by $C_\pi(\iI)$.   The geometric realization of this coset poset has dimension determined by the longest chain of subgroups in $\iI$. In both examples a longest chain looks like
$$
\set{0}=I_0 \to I_1 \to I_2
$$
where $I_i$ are isotropic of dimension $i$. So the resulting space has dimension $2$. Moreover, from the homotopy exact sequence of $\varphi$ it turns out that the $\pi_1(C_\pi(\iI))$ is trivial. As a consequence of Hurewicz theorem \cite{Hat02} a two dimensional simply connected space is necessarily a wedge of $2$-spheres. The number of   spheres can be calculated from the Euler characteristic of the coset poset as in \cite[\S 8]{O14}.

Finally the fundamental group $\pi$ can be computed using the presentation given in \ref{eq:presentation}. 
  In both cases we observe that $\pi$ is abelian. This can be seen  from Fig.\ref{eq:presentation}. Each edge corresponds to a generator in the presentation and the $2$-cells give the relations between the generators. For example, in the Mermin square, the   union of all the $2$-cells of the torus gives the relation $[e_{z_2},e_{x_2}]=1$. Then it follows that the natural map $\pi\to V_2$ is an isomorphism. The element $(e_{z_2})(e_{x_2})$ maps to $z_2+x_2$. We can argue similarly for Mermin's star and conclude that $\pi\cong V_3$ in this case.
}

The Mermin star and square  belong to a class of examples that can be described using the fundamental group of $\Bcx\iI$. Let $\pi(\iI)$ denote the fundamental group $\pi_1(\Bcx\iI)$. The inclusion $\iI\subset \iI(V)$ induces a group homomorphism
$$
\tau:\pi(\iI)\to V\times \ZZ/2.
$$
Let $W$ denote the image, and $K$ denote the kernel of this homomorphism. We have an exact sequence of groups
$$
1\to K \to \pi(\iI) \to W \to 0.
$$
Suppose that $\iI$ contains more than one maximal object, so that $W$ is larger than an isotropic subspace. Then Corollary \ref{cor:Q-beta} implies that the restriction of $\beta$ to $H^*(W)$ is non-zero. The $5$-term exact sequence  associated to   the exact sequence above gives us
$$
0\to H^1(W) \to H^1(\pi(\iI)) \to H^1(K)^W \stackrel{\delta}{\to} H^2(W) \to H^2(\pi(\iI)).
$$
By exactness $\tau^*[\beta]$ is non-zero if it is not in the image of the connecting homomorphism $\delta$. This happens when $K=1$, in which case $\tau$ is injective. So for a cover $\iI$ of contexts that contains more than one maximal objects if the natural map $\pi(\iI)\to V\times \ZZ/2$ is injective then $\eE_\beta(\iI)=\emptyset$. As observed in Proposition \ref{pro:merfibseq} Mermin square and star are both examples of this kind.

\subsection{{\color{black}Mermin's inequality}} 
Using the representation-theoretic language of \S \ref{sec:Wigner} we describe a generalized version of {\color{black}a Bell-type inequality derived from the Mermin star construction.} 

Let $\iI$ be a collection of contexts. Given a context $J$ let us define {\color{black}
\begin{equation}\label{eq:J}
\jJ =  \iI \sqcup \set{I \cap J |\; I\in \iI} \sqcup\set{J}.
\end{equation} }
  The inclusion $f:\iI \to \jJ$ of the collections induces a map between the inverse limits 
$$
f^*:\ilim{\jJ}{} D\eE_\beta \to  \ilim{\iI}{} D\eE_\beta
$$
since a compatible family for $\jJ$ gives a compatible family for the smaller collection $\iI$.

\Pro{\label{pro:CHSH} Let $\iI$ be a collection of contexts {\color{black}with} $\eE_\beta(\iI)\neq \emptyset$. Suppose that $J$ is a context such that the collection  $\jJ$ defined as in \ref{eq:J} satisfies $\eE_\beta(\jJ)=\emptyset$.
Then any common eigenstate $\rho$ of the operators $\set{\eta(a)|\;a\in J}$ is contextual with respect to $\iI$.
}
\Proof{ 
  Let $e_{\rho,\iI}$ and $e_{\rho,\jJ}$ denote the empirical models corresponding to $\iI$ and $\jJ$. 
  Under the natural map $f^*$ induced by the inclusion $\iI\to \jJ$ the empirical model $e_{\rho,\jJ}$ maps to $e_{\rho,\iI}$.
Let ${\color{black}s_0}:J\to {\color{black}\ZZ/p}$ denote the eigenvalue function of $\rho$ {\color{black}($\ZZ/p$ is identified with $\Span{\omega}$ as usual)}. By the orthogonality of eigenstates corresponding to distinct eigenvalue functions we have {\color{black}
$$
e_{\rho,\jJ}|_{I\cap J}(s) = \tr(\rho P_{\chi_s}) = \delta_{s_0|_{I\cap J},s} 
$$ }
for any $I\in \iI$. 
Thus   $e_{\rho,\jJ}$   is determined by $e_{\rho,\iI}$ on isotropic subspaces that belong to $\iI$, and is given by {\color{black}$\delta_{s_0}|_{I\cap J}$} on the rest. Assume that $\rho$ is non-contextual with respect to $\iI$, that is $e_{\rho,\iI}$ can be written as $\theta(d)$ for some $d\in D\eE_\beta(\iI)$. We can write $d= \sum_s p_s[s]$ for some $0\leq p_s\leq 1$, $\sum p_s =1$. 
For a non-zero $p_s$ the function $s$ and $\chi_0$ coincide on each intersection $I\cap J$. Therefore $s$ and {\color{black}$s_0$} together   define an element of $\eE_\beta(\jJ)$, which is a contradiction to the assumption that $\eE_\beta(\jJ)=\emptyset$. 
}  

We can express this as an inequality. Consider the function
$$
{\color{black}\ev_{s_0}: D\eE_\beta(J)} \to [0,1]
$$
{\color{black}defined by  $\ev_{s_0}(\sum p_s [s])= p_{s_0}$.
} 
Assume that there exits $d\in \eE_\beta(\iI)$ such that $e_{\rho,\iI}= \theta(d)$.
Consider a lift $\tilde d$ under the map $f^*$ of the element $\theta(d)${\color{black}, e.g. we can take $\tilde d = e_{\rho,\jJ}$}. Then  $\tilde d|_J$  satisfies  the inequality
\begin{equation}\label{eq:ineq}
{\color{black}\ev_{s_0}}(\tilde d|_J) <1. 
\end{equation}
As in the proof of Proposition \ref{pro:CHSH} this is because $\tilde d|_J$ can be written as a sum {\color{black}$\sum_{s} p_{s} [s]$} with more than one non-zero term, otherwise we would be able to construct an element of $\eE_\beta(\jJ)$ and obtain a contradiction. We can rephrase Proposition \ref{pro:CHSH} by saying that for a fixed {\color{black}$s_0$} every common eigenstate  of $\set{\eta(a)|\;a\in J}$ corresponding to this eigenvalue function violates the inequality, thus these states are contextual.

The typical example is obtained from $\iI_\star$, the cover of contexts for Mermin star, by omitting the fifth context. Let $\iI$ denote the resulting context, and let $J=I_5$. For $\rho$ we take the GHZ state
$$
|\text{GHZ}\rangle = \frac{1}{\sqrt{2}}(|000\rangle +|111\rangle)
$$
which is an eigenstate with eigenvalue function ${\color{black}s_0}:J\to \ZZ/2$ given by $0,1,1$ on the generators $x_1+x_2+x_3$, $y_1+y_2+x_3$, $x_1+y_2+y_3$, respectively. {\color{black} Then \ref{eq:ineq} gives
$$
\begin{aligned}
\ev_{s_0}(\tilde d|_J) &=  \frac{1}{|J|} \sum_{a\in J} (-1)^{s_0(a)} \tr(\rho T_a)  \\
&=\frac{1}{4}\left( \tr(\rho T_{x_1+x_2+x_3})-\tr(\rho T_{x_1+y_2+y_3})-\tr(\rho T_{y_1+x_2+y_3})-\tr(\rho T_{y_1+y_2+x_3})   \right) <1
\end{aligned}
$$ 
which is violated since the left-hand side is equal to $1$.
For non-contextual $\rho$ it can be shown that the left-hand side is $\leq 1/2$.}
This inequality is usually referred to as {\color{black}Mermin's inequality \cite{mermin1990extreme}} in the physics literature{\color{black}; see also \cite{cabello2002bell,Fraction2018}. 
Similarly we can obtain an inequality for the Mermin square construction by taking $J=I_2$. The corresponding inequality $(\tr(\rho T_{x_1+x_2})+\tr(\rho T_{z_1+z_2})-\tr(\rho T_{y_1+y_2}) )/3<1$ is violated for the Bell state $(\bra{00} + \bra{11})/\sqrt{2}$.}


\begin{thebibliography}{10}

\bibitem{aasnaess2020cohomology}
S.~Aasn{\ae}ss.
\newblock Cohomology and the algebraic structure of contextuality in
  measurement based quantum computation.
\newblock {\em arXiv preprint arXiv:2005.00213}, 2020.

\bibitem{abramsky2019comonadic}
S.~Abramsky, R.~S. Barbosa, M.~Karvonen, and S.~Mansfield.
\newblock A comonadic view of simulation and quantum resources.
\newblock In {\em 2019 34th Annual ACM/IEEE Symposium on Logic in Computer
  Science (LICS)}, pages 1--12. IEEE, 2019.

\bibitem{abramsky2015contextuality}
S.~Abramsky, R.~S. Barbosa, K.~Kishida, R.~Lal, and S.~Mansfield.
\newblock Contextuality, cohomology and paradox.
\newblock {\em 24th EACSL Annual Conference on Computer Science Logic},
  41:211--228, 2015.

\bibitem{abramsky}
S.~Abramsky and A.~Brandenburger.
\newblock The sheaf-theoretic structure of non-locality and contextuality.
\newblock {\em New Journal of Physics}, 13(11):113036, 2011.

\bibitem{abramsky2014operational}
S.~Abramsky and A.~Brandenburger.
\newblock An operational interpretation of negative probabilities and
  no-signalling models.
\newblock In {\em Horizons of the mind. A tribute to Prakash Panangaden}, pages
  59--75. Springer, 2014.

\bibitem{abramsky2011cohomology}
S.~Abramsky, S.~Mansfield, and R.~S. Barbosa.
\newblock The cohomology of non-locality and contextuality.
\newblock {\em Electronic Proceedings in Theoretical Computer Science (EPTCS)},
  95:1--14, 2011.

\bibitem{Acin}
A.~Ac{\'\i}n, T.~Fritz, A.~Leverrier, and A.~B. Sainz.
\newblock A combinatorial approach to nonlocality and contextuality.
\newblock {\em Communications in Mathematical Physics}, 334(2):533--628, 2015.

\bibitem{ACT12}
A.~Adem, F.~R. Cohen, and E.~Torres~Giese.
\newblock Commuting elements, simplicial spaces and filtrations of classifying
  spaces.
\newblock {\em Mathematical Proceedings of the Cambridge Philosophical
  Society}, 152(1):91--114, 2012.

\bibitem{AG15}
A.~Adem and J.~M. G\'omez.
\newblock A classifying space for commutativity in {L}ie groups.
\newblock {\em Algebraic \& Geometric Topology}, 15(1):493--535, 2015.

\bibitem{adem2013cohomology}
A.~Adem and R.~J. Milgram.
\newblock {\em Cohomology of finite groups}, volume 309.
\newblock Springer Science \& Business Media, 2013.

\bibitem{AB09}
J.~Anders and D.~E. Browne.
\newblock Computational power of correlations.
\newblock {\em Physical Review Letters}, 102(5):050502, 2009.

\bibitem{As86}
M.~Aschbacher.
\newblock {\em Finite group theory}, volume~10 of {\em Cambridge Studies in
  Advanced Mathematics}.
\newblock Cambridge University Press, Cambridge, 1986.

\bibitem{AS69}
M.~F. Atiyah and G.~B. Segal.
\newblock Equivariant {$K$}-theory and completion.
\newblock {\em J. Differential Geometry}, 3:1--18, 1969.

\bibitem{bell64}
J.~S. Bell.
\newblock On the {E}instein {P}odolsky {R}osen paradox.
\newblock {\em Physics Physique Fizika}, 1(3):195, 1964.

\bibitem{bell1966problem}
J.~S. Bell.
\newblock On the problem of hidden variables in quantum mechanics.
\newblock {\em Reviews of Modern Physics}, 38(3):447, 1966.

\bibitem{BDBOR17}
J.~Bermejo-Vega, N.~Delfosse, D.~E. Browne, C.~Okay, and R.~Raussendorf.
\newblock Contextuality as a resource for qubit quantum computation.
\newblock {\em Physical Review Letters}, 119(120505), 2016.

\bibitem{magic}
S.~Bravyi and A.~Kitaev.
\newblock Universal quantum computation with ideal clifford gates and noisy
  ancillas.
\newblock {\em Physical Review A}, 71(2):022316, 2005.

\bibitem{Brown12}
K.~S. Brown.
\newblock {\em Cohomology of groups}, volume~87.
\newblock Springer Science \& Business Media, 2012.

\bibitem{cabello2002bell}
A.~Cabello.
\newblock Bell’s theorem with and without inequalities for the three-qubit
  greenberger-horne-zeilinger and w states.
\newblock {\em Physical Review A}, 65(3):032108, 2002.

\bibitem{Cabello}
A.~Cabello, S.~Severini, and A.~Winter.
\newblock ({N}on-)contextuality of physical theories as an axiom.
\newblock {\em arXiv preprint arXiv:1010.2163}, 2010.

\bibitem{Chu15}
C.~Cheng.
\newblock A character theory for projective representations of finite groups.
\newblock {\em Linear Algebra and its Applications}, 469:230--242, 2015.

\bibitem{Silva18}
N.~de~Silva and R.~S. Barbosa.
\newblock Contextuality and noncommutative geometry in quantum mechanics.
\newblock {\em Communications in Mathematical Physics}, 365(2):375--429, 2019.

\bibitem{DBGR15}
N.~Delfosse, P.~A. Guerin, J.~Bian, and R.~Raussendorf.
\newblock Wigner function negativity and contextuality in quantum computation
  on rebits.
\newblock {\em Physical Review X}, 5(2):021003, 2015.

\bibitem{Delfosse17}
N.~Delfosse, C.~Okay, J.~Bermejo-Vega, D.~E. Browne, and R.~Raussendorf.
\newblock Equivalence between contextuality and negativity of the {W}igner
  function for qudits.
\newblock {\em New Journal of Physics}, 19(12):123024, 2017.

\bibitem{Dugger}
D.~Dugger.
\newblock A primer on homotopy colimits.
\newblock {\em preprint}, 2008.

\bibitem{dwyer2008twisted}
C.~Dwyer.
\newblock Twisted equivariant {K}--theory for proper actions of discrete
  groups.
\newblock {\em K-Theory}, 38(2):95--111, 2008.

\bibitem{Frankel-gauge}
T.~Frankel.
\newblock {\em The geometry of physics: an introduction}.
\newblock Cambridge university press, 2011.

\bibitem{Friedman12}
G.~Friedman.
\newblock Survey article: an elementary illustrated introduction to simplicial
  sets.
\newblock {\em The Rocky Mountain Journal of Mathematics}, pages 353--423,
  2012.

\bibitem{Gro06}
D.~Gross.
\newblock Hudson’s theorem for finite-dimensional quantum systems.
\newblock {\em Journal of mathematical physics}, 47(12):122107, 2006.

\bibitem{Hat02}
A.~Hatcher.
\newblock {\em Algebraic topology}.
\newblock Cambridge University Press, Cambridge, 2002.

\bibitem{Howard14}
M.~Howard, J.~Wallman, V.~Veitch, and J.~Emerson.
\newblock Contextuality supplies the ‘magic’ for quantum computation.
\newblock {\em Nature}, 510(7505):351, 2014.

\bibitem{kac2000vacuum}
V.~G. Kac and A.~V. Smilga.
\newblock Vacuum structure in supersymmetric {Y}ang--{M}ills theories with any
  gauge group.
\newblock In {\em The Many Faces of the Superworld: Yuri Golfand Memorial
  Volume}, pages 185--234. World Scientific, 2000.

\bibitem{kochen}
S.~Kochen and E.~P. Specker.
\newblock The problem of hidden variables in quantum mechanics.
\newblock In {\em The logico-algebraic approach to quantum mechanics}, pages
  293--328. Springer, 1975.

\bibitem{Lah12}
A.~Lahtinen.
\newblock The {A}tiyah--{S}egal completion theorem in twisted {K}--theory.
\newblock {\em Algebraic \& Geometric Topology}, 12(4):1925--1940, 2012.

\bibitem{mermin1990extreme}
N.~D. Mermin.
\newblock Extreme quantum entanglement in a superposition of macroscopically
  distinct states.
\newblock {\em Physical Review Letters}, 65(15):1838, 1990.

\bibitem{Mer93}
N.~D. Mermin.
\newblock Hidden variables and the two theorems of {J}ohn {B}ell.
\newblock {\em Reviews of Modern Physics}, 65(3):803, 1993.

\bibitem{Milnor-classifying}
J.~Milnor.
\newblock Construction of universal bundles, {I}.
\newblock {\em Annals of Mathematics}, pages 272--284, 1956.

\bibitem{Nielsen02}
M.~A. Nielsen and I.~Chuang.
\newblock {\em Quantum computation and quantum information}.
\newblock Cambridge: Cambridge University Press, 2002.

\bibitem{O14}
C.~Okay.
\newblock Homotopy colimits of classifying spaces of abelian subgroups of a
  finite group.
\newblock {\em Algebraic \& Geometric Topology}, 14(4):2223--2257, 2014.

\bibitem{O15}
C.~Okay.
\newblock Colimits of abelian groups.
\newblock {\em J. Algebra}, 443:1--12, 2015.

\bibitem{O16}
C.~Okay.
\newblock Spherical posets from commuting elements.
\newblock {\em Journal of Group Theory}, 21(4):593--628, 2018.

\bibitem{Coho}
C.~Okay, S.~Roberts, S.~Bartlett, and R.~Raussendorf.
\newblock Topological proofs of contextuality in quantum mechanics.
\newblock {\em Quantum Information and Computation 17, 1135-1166 (2017)}.

\bibitem{Fraction2018}
C.~Okay, E.~Tyhurst, and R.~Raussendorf.
\newblock The cohomological and the resource-theoretic perspective on quantum
  contextuality: common ground through the contextual fraction.
\newblock {\em Quantum Information and Computation 18, 1272-1294 (2018)}.

\bibitem{Q71a}
D.~Quillen.
\newblock The {${\rm mod}$} {$2$} cohomology rings of extra-special
  {$2$}-groups and the spinor groups.
\newblock {\em Mathematische Annalen}, 194:197--212, 1971.

\bibitem{MBQC}
R.~Raussendorf and H.~J. Briegel.
\newblock A one-way quantum computer.
\newblock {\em Physical Review Letters}, 86(22):5188, 2001.

\bibitem{RBDOB17}
R.~Raussendorf, D.~E. Browne, N.~Delfosse, C.~Okay, and J.~Bermejo-Vega.
\newblock Contextuality and {W}igner function negativity in qubit quantum
  computation.
\newblock {\em Physical Review A}, 95(5):052334, 2017.

\bibitem{serre1977linear}
J.-P. Serre.
\newblock {\em Linear representations of finite groups}.
\newblock Springer, 1977.

\bibitem{Spekkens}
R.~W. Spekkens.
\newblock Contextuality for preparations, transformations, and unsharp
  measurements.
\newblock {\em Physical Review A}, 71(5):052108, 2005.

\bibitem{Steenrod}
N.~E. Steenrod.
\newblock {\em The topology of fibre bundles}, volume~14.
\newblock Princeton university press, 1999.

\bibitem{weibel1995introduction}
C.~A. Weibel.
\newblock {\em An introduction to homological algebra}.
\newblock Number~38. Cambridge university press, 1995.

\bibitem{witten1982constraints}
E.~Witten.
\newblock Constraints on supersymmetry breaking.
\newblock {\em Nuclear Physics B}, 202(2):253--316, 1982.

\end{thebibliography}

\end{document}